\documentclass[cmp]{svjour}
\usepackage{times}
\usepackage{natbib}
\usepackage{amsmath}
\usepackage{theorem}
\usepackage{amscd}
\usepackage{amsfonts,amssymb}
\usepackage[hypertex]{hyperref}
\def\urlprefix\url#1{\url{#1}} 
\providecommand{\url}[1]{\texttt{#1}}

\theoremstyle{plain}
\newtheorem{defiprop}[definition]{Definition and proposition}
\smartqed
\let\oendproof\endproof
\def\endproof{\qed\oendproof}

\newcommand{\C}{{\mathbb{C}}}
\newcommand{\N}{{\mathbb{N}}}
\newcommand{\R}{{\mathbb{R}}}
\newcommand{\Z}{{\mathbb{Z}}}
\newcommand{\ONE}{{\boldsymbol{1}}}
\newcommand{\CS}{{$C^*$}}
\newcommand{\WS}{{$W^*$}}
\newcommand{\hookllongrightarrow}{\lhook\joinrel\relbar\joinrel\longrightarrow}
\def\<#1|#2>{\left\langle #1\vphantom{#2}\right|\left.\negmedspace\vphantom{#1}#2\right\rangle}

\DeclareMathOperator{\tr}{tr}

\DeclareMathOperator{\spec}{spec}
\DeclareMathOperator{\supp}{supp}
\DeclareMathOperator{\RR}{RR}
\DeclareMathOperator{\dist}{dist}

\DeclareMathOperator{\ran}{ran}

\DeclareMathOperator{\End}{End}
\def\ma{{\underline{*}}}

\begin{document}
\title{Non-commutative Bloch Theory}
\author{Michael J.\ Gruber}
\institute{
Department of Mathematics, MIT 2--167,
77 Massachusetts Avenue,
Cambridge, MA 02139--4307,
USA\\
\email{mjg@math.mit.edu}
}
\date{\today}
\maketitle
\begin{abstract}
For differential operators which are invariant under the 
action of an abelian group Bloch theory is the preferred
tool to analyze spectral properties. By shedding some 
new non-commutative light on this we motivate
the introduction of a non-commu\-ta\-tive Bloch theory for elliptic 
operators on Hilbert C$^*$-modules. It relates
properties of C$^*$-algebras to spectral properties of module
operators such as band structure, weak genericity
of cantor spectra, and absence of discrete spectrum. It 
applies e.g.\ to differential operators invariant under 
a projective group action, such as Schr\"odinger, Dirac and Pauli
 operators with periodic magnetic field, as well as to discrete models,
such as the almost Matthieu equation and the quantum pendulum.

\end{abstract}

\section{Introduction}
Bloch (or Floquet) theory in its usual form has a long history already.
Basically it starts from the fact that partial differential equations
with
constant coefficients are mapped into algebraic equations by means of
the
Fourier or Laplace transform.
Now, if the coefficients are not constant but just periodic under an
abelian
(locally compact topological) group one still has the Fourier transform
on
such groups, mapping functions on the group $\Gamma$ into functions on
the
dual group $\hat\Gamma$;
the original spectral problem on a non-compact manifold is mapped into
a
(continuous) sum of spectral problems on a compact manifold (see
Section
\ref{sec:CBT}).

This is what makes Bloch theory an indispensable tool especially for
solid
state physics, where one describes the motion of non-interacting
electrons
in a periodic solid crystal by a Schr\"odinger operator
$ -\Delta + V$ on $L^2(\R^d)$.
The potential function $V$ is the gross electric potential generated by
all
the crystal ions and thus is periodic under the lattice given by the
crystal
symmetry. Bloch theory shows that the spectrum of the periodic Schr\"odinger operator
 has band structure in the following sense:
\begin{definition}[band structure] \label{definition:band structure}
A subset of the real line has band structure if it is a locally finite union of
closed intervals.
\end{definition}
Band structure is an essential ingredient of electronic transport in metals and semi-conductors.
By exploiting Bloch theory and the structure of the Schr\"odinger operator further one can see
that the spectrum is purely absolutely continuous, which is sometimes included in the definition of band structure.

Measurements of crystals often require magnetic fields $b$ (2-form).
In quantum mechanics, they are described by a vector potential (1-form)
$a$
such that $b=da$ ($B=\operatorname{curl}A$ for the corresponding vector
fields).
The magnetic Schr\"odinger operator
then reads \[ H=-(\nabla-ia)^2+V. \]

But, although $b$ is periodic or even constant, $a$ need not be so, and
$H$
won't be periodic.
It is therefore necessary to use magnetic translations (first
introduced by
 \citet{Zak:DESEF}) under which $H$ still is invariant.
But now, these translations do not commute with each other in general.
Therefore ordinary (commutative) Bloch theory does not apply.

Basically, the reason for this failure is that a non-abelian group has
no
``good'' group dual: the set of (equivalence classes of) irreducible
representations has no natural group structure whereas the set of
one-dimensional representations is too small to describe the group ---
otherwise it would be abelian.

But although $\hat\Gamma$ does not exist any more, the algebra
$C(\hat\Gamma)$
 of continuous functions continues to exist in some sense:
It is given by the reduced group \CS-algebra of $\Gamma$ which is just
the
\CS-algebra generated by $\Gamma$ in its regular representation on
itself
(on $l^2(\Gamma)$).

Section \ref{sec:CBTNPV} shows how one can re-formulate ordinary Bloch
theory in
a way which refrains from using the points of $\hat\Gamma$ and relies
just on
 the r\^ole of $C(\hat\Gamma)$.
{}From a technical point of view this requires switching from
measurable fields
 of Hilbert spaces to continuous fields which then can be described as
Hilbert
\CS-modules over the commutative \CS-algebra $C(\hat\Gamma)$.

Having done this one can retain the setup but omit the condition of
commutativity for the \CS-algebra $C(\hat\Gamma)$.
Thus one is lead to non-commutative Bloch theory (Section
\ref{sec:NBT}) dealing
 with elliptic operators on Hilbert \CS-modules over non-commutative
\CS-algebras.
The basic task is now to relate properties of the \CS-algebra to
spectral
properties of ``periodic'' operators.
Thus one generalizes spectral results for elliptic operators on compact
manifolds as well as  results of ordinary Bloch theory:
\begin{theorem}
Isolated eigenvalues of $\mathcal A$-elliptic operators  have $\mathcal A$-finite eigenprojections, their eigenspaces
have finite $\tau$-dimension.

Under certain assumptions they have essential spectrum only (isolated eigenvalues have infinite multiplicity).
\end{theorem}
See Theorem~\ref{theorem:isolated eigenvalue} for exact assumptions (they are fulfilled by
Schr\"odinger operators with periodic magnetic field).

Non-commutative Bloch theory allows to treat continuous and discrete models, i.e.\ differential
and difference operators, on equal footing. It opens the way to apply a
result of \cite{ChoEll:DSAEFSIRCA} on weak genericity of Cantor spectra
in discrete models to the continuous models also, i.e.\ to the phenomenon
opposite to band structure:
\begin{definition}[Cantor set]
A Cantor set is a subset of a topological space which is no\-where
dense
 (the closure has empty interior) and has no isolated points.
\end{definition}
Now the \CS-algebras of symmetries determines which of the two opposite spectral types is present:
\begin{definition}[Kadison property]\label{definition:Kadison property}
The \emph{Kadison constant} K of a \CS-algebra $\cal A$ together with a trace $\tau$ is defined by
\begin{align}
K = \inf\{\tau(P)\mid 0\ne P\in\mathcal{A}\text{ projection}\}.
\end{align}
We say the pair $(\mathcal{A},\tau)$ has the \emph{Kadison property} if $K>0$.
\end{definition}
\begin{theorem}[band structure]
If $(\mathcal{A},\tau)$ has the Kadison property, then the spectrum of every
symmetric $\mathcal{A}$-elliptic operator has band structure.
\end{theorem}
(See Theorem~\ref{theorem:band structure}.) This applies e.g.\ to magnetic Schr\"odinger operators in the case of rational magnetic flux.

Opposite to the Kadison property is the property $RRI_0$ (see Definition~\ref{defiprop:real rank}),
and it is a criterion for the opposite spectral type:
\begin{theorem}[Cantor spectrum]
If $(\mathcal{A},\tau)$ has property $RRI_0$ then every $\mathcal A$-elliptic operator can be approximated
arbitrarily well (in norm resolvent sense) by one which has Cantor spectrum.
\end{theorem}
(See Theorem~\ref{theorem:Cantor spectrum}.) The important issue here is that the approximation takes place within a natural \CS-algebra generated by
symmetries connected to the operator. Approximation within a von Neumann algebra would be pointless, of course.
This theorem applies e.g.\ to magnetic Schr\"odinger operators on $\R^2$ in the case of irrational flux.

In Section \ref{sec:A} we list examples where non-commutative Bloch
theory applies: gauge-periodic elliptic differential operators
(Schr\"odinger, Pauli, Dirac with periodic magnetic field) and difference operators (almost
Matthieu, quantum pendulum).

For the convenience of the reader we add an appendix on continuous
fields of Hilbert spaces and on Hilbert \CS-modules and their GNS
representation.

\bigskip
A short overview of this paper appeared in \cite{Gru:NBTO}.
\nocite{Gru:NB}

\begin{acknowledgements}
I am indebted to my thesis advisor Jochen Br\"uning for his scientific
support.

This work has been supported financially by Deutsche
Forschungsgemeinschaft
(DFG) as project D6 at the SFB 288 (differential geometry and quantum
physics), Berlin.
\end{acknowledgements}


\section{Commutative Bloch theory} \label{sec:CBT}
In this section we recall the basic elements of Bloch theory for
periodic operators in the geometric context of  vector bundles, since
even in the scalar case of a magnetic Schr\"odinger operator
one is lead to consider
 possibly non-trivial complex line bundles.
The standard reference for the theory of direct integrals is
\cite[chapter II]{Dix:AOEHAN}, for Bloch theory in Euclidean space see
\cite[chapter XIII.16]{ReeSim:AO}.

Our general assumptions are: $X$ is an oriented smooth Riemannian
manifold without boundary, $\Gamma$ a discrete abelian group acting on
$X$ freely, isometrically, and properly discontinuously.
Furthermore, we assume the action to be cocompact in the sense that the
quotient $M:=X/\Gamma$ is compact.

Next, let $E$ be a smooth Hermitian vector bundle over $X$.

\begin{example}[solid crystals]
The main motivating example for our setting comes from solid state
physics.
Here, $X=\R^n$ is the configuration space of a single electron
($n=2,3$).
It is supposed to move in a crystal whose translational symmetries are
described by a lattice $\Z^n\simeq\Gamma\subset\R^n$, which acts on $X$
by translations, of course.
Note that this does not take into account the point symmetries.
$\Gamma$ could be extended by them but the action would not be free any
more.
Considering just the translations is enough to achieve the compactness
of the quotient $M\simeq T^n.$

Wave functions of electrons are just complex-valued functions on
$\R^n$, so we can set $E=\R^n\times\C$.
One may also include the spin of the electrons into the picture by
choosing the appropriate trivial spinor bundle $E=\R^n\times\C^k$.
\end{example}

\begin{definition}[periodic operator]\label{definition:periodic operator}
Assume there is an isometric  lift $\gamma_*$ of the action of $\gamma$
from $X$ to $E$ in the following sense:
\begin{align}
\gamma_*:E_x\rightarrow E_{\gamma x}\text{ for }x\in X,\gamma\in\Gamma.
\end{align}
This defines an action $T_\gamma$ on the sections: For $s\in
C^\infty_c(E)$ we define
\begin{align}
(T_\gamma s)(x) := \gamma_* s(\gamma^{-1}x)\text{ for }x\in
X,\gamma\in\Gamma. \label{equ:actiononsections}
\end{align}
$(T_\gamma)_{\gamma\in\Gamma}$ induces a unitary representation of
$\Gamma$ in $L^2(E)$ since $\gamma_*$ acts isometrically and
$T_\gamma^*=(T_\gamma)^{-1}$.

A differential operator $D$ on ${\mathcal D}(D):=C^\infty_c(E)$ is
called periodic if, on $\mathcal{D}(D)$, we have:
\begin{equation} \forall{\gamma\in\Gamma}: [T_\gamma,D]=0
\end{equation}
\end{definition}

\begin{example}[periodic Schr\"odinger operator]
Given a manifold as described above, we may lift the action to any
trivial vector bundle $E:=X\times\C^k$ canonically.
If $D$ is a periodic operator on $X$ (for example any geometric
operator, i.e.\ defined by the metric on $X$) and $V\in
C^\infty(X,M(k,\C))$ a periodic field of endomorphisms, then $D+V$ is a
periodic operator on $E$.

In the case of a crystal, we choose the Laplacian (which describes the
kinetic energy quantum mechanically) and a periodic potential $V\in
C^\infty(\R^n,\R)$ (which describes the electric field of the ions at
the lattice sites) to get the periodic Schr\"odinger operator
$\Delta+V$.
\end{example}

\begin{example}[Schr\"odinger operator with exact periodic magnetic
field]\label{example:sowepmf}
Let $b\in\Omega^2(X)$ be a magnetic field 2-form.
In dimension 3 this corresponds (by the Hodge star) to a vector field
$B$, in dimension 2 to a scalar function which may be thought of as the
length and orientation of a normal vector $B$.
From physical reasons one has $\operatorname{div} B=0$, i.e.\ $db=0$.
For simplicity we assume that $b$ is not only closed but exact, so
there is $a\in\Omega^1(X)$ with $b=da$ ($B=\operatorname{rot} A$ for
the corresponding vector fields).
This defines a magnetic Hamiltonian operator
\begin{align} \Delta^a:=(d-\imath a)^*(d-\imath a) \label{equ:BL}
\end{align}
(the \emph{minimally coupled Hamiltonian}), where $d$ is the ordinary
differential (corresponding to the gradient) and ${}^*$ the  adjoint of
an operator between the Hilbert spaces of $L^2$-functions $L^2(X)$ and
of $L^2$-1-forms $L^2(X,\Lambda T^*X)$.

\def\ma{{\underline{*}}}
For later convenience we set, for $\gamma\in\Gamma$ and
$\omega\in\Omega(X)$, $\gamma^\ma\omega:=\left(\gamma^{-
1}\right)^*\omega$, considering $\gamma^{-1}$ as a map $X\rightarrow X$
and using the usual pull-back of forms. This puts the action on forms
in a notation compatible with the action on
sections~\eqref{equ:actiononsections} from the preceding definition.

Now, if $b$ is periodic, $a$ does not need to be so: If
$b\in\Omega^2(\R^n)$ is constant then $a$ is affine linear. So the
translations are no symmetries for the magnetic Hamiltonian.
\citet{Zak:DESEF} was the first to define the so-called \emph{magnetic
translations}: Since $d(a-\gamma^\ma a)=da-\gamma^\ma da=b-\gamma^\ma
b=0$, one can (at least if $H^1(X)=0$) find a function $\chi_\gamma$
with $d\chi_\gamma=a-\gamma^\ma a$.
One may define such a function  explicitly by
\[ \chi_\gamma(x):=\int_{x_0}^x(a-\gamma^\ma a) \]
which is well-defined if $H_1(X)=0$.
If we now define a gauge function $s_\gamma := e^{\imath\chi_\gamma}$
then
\begin{align*}
  (d-\imath a)(s_\gamma \gamma^\ma f) &= s_\gamma \gamma^\ma df +
\imath(a-\gamma^\ma a)s_\gamma
\gamma^\ma f -\imath a s_\gamma \gamma^\ma f \\
 &= s_\gamma \gamma^\ma df -\imath \gamma^\ma a\gamma^\ma f \\
 &= s_\gamma \gamma^\ma df -\imath s_\gamma \gamma^\ma(a f) \\
 &= s_\gamma \gamma^\ma \left( (d-\imath a) f \right).
\end{align*}
So we have found symmetries of the magnetic Hamiltonian operator, the
gauged translations
\begin{align*}
T_\gamma: C^\infty(X) &\rightarrow C^\infty(X), \\
( T_\gamma s)(x) &= s_\gamma(x) (\gamma^\ma s)(x)
\end{align*}
coming from the lifted action
\begin{align*}
\gamma_*: X\times\C&\rightarrow X\times\C,\\
 \gamma_*(x,c) &= \left(\gamma x, s_\gamma(x)c\right).
\end{align*}
The commutation relation for the magnetic translations is
\begin{align}
(T_{\gamma_1}&T_{\gamma_2}s)(x) =
s_{\gamma_1}(x)s_{\gamma_2}(\gamma_1^{-1}x)s(\gamma_2^{-1}\gamma_1^{-
1}x) \notag \\
&=\exp\left(\imath\left(\int_{x_0}^x a-\gamma_1^\ma
a+\int_{x_0}^{\gamma_1^{-1}x}a-\gamma_2^\ma a \right)\right)
s(\gamma_2^{-1}\gamma_1^{-1}x) \notag \\
&=\exp\left(\imath\left(\int_{x_0}^x a-\gamma_1^\ma
a+\int_{\gamma_1x_0}^{x}\gamma_1^\ma a-(\gamma_1\gamma_2)^\ma a
\right)\right) s(\gamma_2^{-1}\gamma_1^{-1}x) \notag   \\
&=\exp\left(\imath\left(\int_{x_0}^{\gamma_1x_0} (\gamma_1\gamma_2)^\ma
a-\gamma_1^\ma a + \int_{x_0}^x a-(\gamma_1\gamma_2)^\ma
a\right)\right) s(\gamma_2^{-1}\gamma_1^{-1}x)\notag  \\
&=\exp\left(\imath\left(\int_{x_0}^{\gamma_1x_0} (\gamma_1\gamma_2)^\ma
a-\gamma_1^\ma a\right)\right) s_{\gamma_1\gamma_2}(x)s(\gamma_2^{-
1}\gamma_1^{-1}x) \label{equ:flux} \\
&=:\Theta(\gamma_1,\gamma_2)s_{\gamma_1\gamma_2}(x)s(\gamma_2^{-
1}\gamma_1^{-1}x)\notag \\
&=\Theta(\gamma_1,\gamma_2)(T_{\gamma_1\gamma_2}s)(x) \notag
\end{align}
with $\Theta(\gamma_1,\gamma_2)\in S^1$. In general this is just a
projective representation of $\Gamma$.
If $a$ itself  is periodic, then $\chi_\gamma=0$ for $\gamma\in\Gamma$,
i.e.\ there is no gauge, and we have just ordinary translations forming
a proper representation.

But even if $a$ is not periodic it can happen that the magnetic
translations commute with each other. This is called the case of
\emph{integral flux} since the term occurring in the exponential in
line~\eqref{equ:flux} is just the magnetic flux through one lattice
face. A periodic $a$ obviously gives rise to zero magnetic flux.

Furthermore, if $V\in C^\infty(X,\R)$ is $\Gamma$-periodic it commutes
with the magnetic translations as well, so $\Delta^a+V$ is a (symmetric
elliptic) periodic operator.

Finally, the very same magnetic translations can be used for the Pauli
Hamiltonian and the magnetic Dirac operator.
\end{example}
\begin{remark}[integral flux]
In the case of integral flux mentioned above quite opposite spectral
phenomena can occur: Periodic Schr\"odinger operators have absolutely
continuous band spectrum, whereas the Landau Hamiltonian on $\R^2$
(constant magnetic field, no electric potential) exhibits pure point
spectrum of infinite degeneracy. In \cite{Gru:MFSASCSMSO} we show that
these are indeed the only phenomena that can occur (although possibly
combined) in the case of integral flux.
\end{remark}

\begin{remark}[non-integral flux]
If the magnetic flux is rational one can find a super-lattice of
$\Gamma$, i.e.\ a subgroup of finite index, such that the flux is
integral. The quotient will still be compact, of course, so that the
rational case can be completely reduced to the integral.

If the magnetic flux is irrational there is no such super-lattice.
Still, one may try to make use of the projective representation defined
above.
There are several approaches, similar in the objects which are used,
different in the objectives that are aimed at and accordingly in the
results.
Our approach will mimic Bloch theory non-commutatively, see
Section~\ref{sec:CBTNPV}.
\end{remark}

\begin{remark}[non-exact magnetic field]\label{remark:nemf}
If $b$ is closed but not exact one first has to agree upon the
quantization procedure used.
\eqref{equ:BL} may be identified as a Bochner Laplacian for a
connection with curvature $b$, and such a connection exists if and only
if $b$ defines an integral cohomology class, i.e.\ $[b]\in H^2(X,\Z)$.
There may exist different quantizations for the same magnetic field.
This is connected to the Bloch decomposition again. For this and the
construction of the magnetic translations in this case see
\cite{Gru:BTQMS}.
\end{remark}

\begin{lemma}[associated bundle]
$E$ is the lift  $\pi^*E'$ of a Hermitian vector bundle $E'$ over $M$
by the projection $\pi:X\rightarrow M$. $E$ and $X$ are $\Gamma$-
principal fiber bundles over $E'$ resp.\ $M$.

To every $\Gamma$-principal fiber bundle and every character
$\chi\in\hat\Gamma$
we associate a line bundle.
This gives the relations depicted in 
diagram~\ref{fig:PAV}
  (``$\rightsquigarrow$'' denotes association of line bundles.).

\begin{figure}[htbp] 
\begin{equation*}
\begin{CD}
       @.   \C^N @.   \C^N  @.  @. @. \C^N @. \C^N\\
   @.       @VVV      @VVV   @. @. @VVV @VVV  \\
\Gamma &\hookllongrightarrow& E    @>\textstyle\pi_*>> E'    @.
\quad\rightsquigarrow\quad @. \C &\hookllongrightarrow& E_\chi    @>>>
E' \\
   @.       @VV\textstyle\pi^EV      @VV\textstyle\pi^{E'}V @.
@. @VVV @VVV \\
\Gamma &\hookllongrightarrow& X    @>\textstyle\pi>> M   @.
\quad\rightsquigarrow\quad @. \C &\hookllongrightarrow& F_\chi    @>>>
M
\end{CD}
\end{equation*}
\caption{
principal fiber bundles and associated line bundles
\label{fig:PAV}}
\end{figure}

In this situation we have $E_\chi\simeq E'\otimes F_\chi$.
\end{lemma}
\begin{proof}
$E$ is a $\Gamma$-principal fiber bundle, so we can use the lifted
$\Gamma$-action  to define $E':=E/\Gamma$.
Since this action is a lift of the $\Gamma$-action on $X$, $E'$ has a
natural structure of a vector bundle over $M$. If
$\pi^{E'}:E'\rightarrow M$ is the bundle projection of $E'$, then the
pull back by $\pi$  is defined as
\begin{align*}
\pi^* E' &= X\times_\pi E'=\{ (x,e)\in X\times E'\mid
\pi(x)=\pi^{E'}(e) \}.
\end{align*}
If $\pi^{E}:E\rightarrow X$ is the bundle projection of $E$ and
$\pi_*:E\rightarrow E'$ is the quotient map, then we get a bundle
isomorphism $E\rightarrow \pi^* E'$  by
\begin{align*}
E\ni e \mapsto (\pi^E(e),\pi_*(e)) \in \pi^* E'.
\end{align*}
Therefore, in this representation the lift $\gamma_*$ of $\gamma$ acts
on $(x,e)\in\pi^*E'$ as $\gamma_*(x,e)=(\gamma x,e)$.

Sections into an associated bundle $P\times_\rho V$ are just those
sections of the bundle $P\times V$ which have the appropriate
transformation property.
By construction, $E_\chi$ is a complex line bundle over $E'$, but from
$E$ it inherits the vector bundle structure, so its sections fulfill:
\begin{equation} C^\infty(E_\chi) \simeq
C^\infty(E)^{\Gamma,\chi}=\{s\in C^\infty(E)\mid
\forall{\gamma\in\Gamma}:\gamma^*s=\chi(\gamma)s \}
\label{equ:Echi=EGchi} \end{equation}
An analogous equation holds for the line bundle $F_\chi$ over $M$.
Finally, \eqref{equ:Echi=EGchi} shows
\begin{align*} E_\chi &= E\times_\chi\C \\
 &= (\pi^* E')\times_\chi\C \\
 &= (X\times_\pi E')\times_\chi\C \\
 &\simeq E'\otimes (X\times_\chi\C) \\
 &= E'\otimes F_\chi.
\end{align*}
Here, all equalities are immediate from the definitions, besides the
last but one, which may be seen as follows:
\begin{align*}
(X\times_\pi E')\times_\chi\C &= (X\times_\pi E'\times\C)/\Gamma
\intertext{with the $\Gamma$-action}
\gamma(x,e,z) &= (\gamma x,e,\chi(\gamma)z),
\intertext{whereas}
E'\otimes (X\times_\chi\C) &=E'\otimes ((X\times\C)/\Gamma)
\intertext{with the $\Gamma$-action}
\gamma(x,z) &= (\gamma x,\chi(\gamma) z).
\end{align*}
So, both bundles are quotients of isomorphic bundles with respect to
the same $\Gamma$-action.
 \end{proof}

\begin{example}[magnetic bundles]
Consider again the case of the magnetic translations for a periodic
magnetic 2-form $b\in\Omega^2(X)$, $E$ being a complex line bundle with
curvature $b$ ($b\in H^2(X,\Z)$). Hence we have $c_1(E)=[b]$ for the
Chern class (up to factors of $2\pi$, depending on the convention).
Since $b$ is periodic we may restrict it to a form $b_M\in\Omega^2(M)$
on the quotient.
The existence of the lifted action, i.e.\ the fact that $E$ can be
written as a pull-back $E=\pi^*E'$, corresponds to the integrality of
$b_M$ from $c_1(E')=[b_M]\in H^2(M,\Z)$.
Tensoring $E'$ with the flat line bundle $F_\chi$ does not change the
Chern class (up to torsion).
In particular, in dimension 2 the integrality of $b_M$ is equivalent to
the integrality of the flux, and $E'$ is trivial only for zero flux.
\end{example}

Next we want to decompose the Hilbert space $L^2(E)$ of square-
integrable sections of $E$ into a direct integral over the character
space $\hat\Gamma$.
On $\hat\Gamma$ we use the Haar measure.
From the theory of representations of locally compact groups we need
the following character relations for abelian discrete $\Gamma$, i.e.\
for abelian, compact $\hat\Gamma$ \cite[see e.g.\ ][ \S 1.5]{Rud:FAG}:

\begin{lemma}[character relations]
For $\gamma\in\Gamma$
\begin{equation} \int_{\hat\Gamma}\chi(\gamma)\,d\chi = \left\{
\begin{array}{l} 1, \quad \gamma=e, \\ 0, \quad \gamma\neq e.
\end{array} \right. \label{equ:Charakter}\end{equation}
For $\chi,\chi'\in\hat\Gamma$
\begin{equation} \sum_{\gamma\in\Gamma} \bar\chi(\gamma)\chi'(\gamma) =
\delta(\chi-\chi') \label{equ:Charakter'}\end{equation}
in distributional sense, i.e.\ for $f\in C(\hat\Gamma)$
\[ \sum_{\gamma\in\Gamma}
\int_{\hat\Gamma}\bar\chi(\gamma)\chi'(\gamma)f(\chi)\,d\chi =
f(\chi'). \]
\end{lemma}

We define for every character $\chi\in\hat\Gamma$ a mapping
$\Phi_\chi:C^\infty_c(E)\ni s\mapsto \tilde s_\chi\in C^\infty(E)$ by
\begin{equation} \tilde s_\chi(x) := \sum_{\gamma\in\Gamma}
\chi(\gamma)\gamma_*s(\gamma^{-1}x). \label{equ:schi} \end{equation}
Since
\begin{align*}
 \tilde s_\chi(\gamma' x) &=
\sum_{\gamma\in\Gamma}\chi(\gamma)\gamma_*s(\gamma^{-1}\gamma'x) \\
 &= \sum_{\gamma\in\Gamma}\chi(\gamma'\gamma'{}^{-
1}\gamma)(\gamma'\gamma'{}^{-1}\gamma)_*s\left((\gamma'{}^{-
1}\gamma)^{-1}x\right) \\
 &= \chi(\gamma')\gamma'_* \tilde s_\chi(x)
\end{align*}
we have
\[ \tilde s_\chi\in C^\infty(E)^{\Gamma,\chi}=\{r\in C^\infty(E)\mid
\forall_{\gamma\in\Gamma} T_\gamma r = \chi(\gamma) r\} \]
which defines a section $s_\chi\in C^\infty(E_\chi)$.

Let $\mathcal D$ be a fundamental domain for the $\Gamma$-action, i.e.\
an open subset of $X$ such that
$\bigcup_{\gamma\in\Gamma}{\gamma\mathcal D}=X$ up to a set of measure
0 and $\gamma\mathcal D\cap\mathcal D=\emptyset$ for $\gamma\not=e$.
Then
\begin{align*}
 \int_{\hat\Gamma} \|s_\chi\|^2_{L^2(E_\chi)}d\chi &= \int_{\hat\Gamma}
\int_{\mathcal{D}} |\tilde s_\chi(x) |^2 dx\,d\chi \\
 &= \int_{\mathcal{D}} \int_{\hat\Gamma}
\sum_{\gamma_1,\gamma_2\in\Gamma} \chi(\gamma_1^{-1}\gamma_2)\langle
{\gamma_1}_*s(\gamma_1^{-1}x)\mid{\gamma_2}_*s(\gamma_2^{-
1}x)\rangle_{E} d\chi\, dx \\
 &= \int_{\mathcal{D}} \sum_{\gamma\in\Gamma} |s(\gamma^{-1}x)|^2 dx \\
 &= \|s\|^2_{L^2(E)}.
\end{align*}
On the one hand, this shows that we can define a measurable structure
on $\prod_{\chi\in\hat\Gamma}L^2(E_\chi)$ by choosing a sequence in
$C^\infty_c(E)$ which is total in  $L^2(E)$.
On the other hand, we can see that the direct integral
$\int^\oplus_{\hat\Gamma}L^2(E_\chi)\,d\chi$ is isomorphic to $L^2(E)$
via the isometry
$\Phi$, whose inverse is given by
\[ \Phi^*\colon(s_\chi)_{\chi\in\hat\Gamma} \mapsto \int_{\hat\Gamma}
\tilde s_\chi(x)\,d\chi, \]
as is easily seen from the character relations \eqref{equ:Charakter}
and \eqref{equ:Charakter'}.

This shows

\begin{lemma}[direct integral]
 The mapping defined by \eqref{equ:schi} can be extended continuously
to a unitary
\begin{equation} \Phi: L^2(E) \rightarrow \int^\oplus_{\hat\Gamma}
L^2(E_\chi)\,d\chi. \end{equation}
\end{lemma}

For the direct integral of Hilbert spaces
$H=\int^\oplus_{\hat\Gamma}H_\chi d\chi$
the set of decomposable bounded operators
$L^\infty(\hat\Gamma,{\mathcal L}(H))$  is given by the commutant
$(L^\infty(\hat\Gamma,\C))'$ in ${\mathcal L}(H)$.
Since commutants are weakly closed and $C(\hat\Gamma,\C)$ is weakly
dense in $L^\infty(\hat\Gamma,\C)$ one has
$(L^\infty(\hat\Gamma,\C))'=(C(\hat\Gamma,\C))'$.
Therefore, in order to determine the decomposable  operators one has to
determine the action of $C(\hat\Gamma)$ on $L^2(E)$. This is easily
done using the explicit form of $\Phi$:

\begin{proposition}[$C(\protect\hat{\Gamma})$-
action]\label{proposition:faction}
$f\in C(\hat\Gamma)$ acts on $s\in C^\infty_c(E)$ by
\begin{align}
 M_fs &:= \Phi^*f\Phi s, \\
\intertext{and one has}
 (M_fs)(x) &= \sum_{\gamma\in\Gamma} \hat f(\gamma^{-1})T_\gamma
s(x),\text{ where} \label{equ:faction} \\
 \hat f(\gamma) &:= \int_{\hat\Gamma}f(\chi)\bar\chi(\gamma)\,d\chi
\label{equ:fhat=intf}
\end{align}
is the Fourier transform of $f$.
$M_f$ is a bounded operator with norm $\|f\|_\infty$.
\end{proposition}
\begin{proof}
For $x\in X$ one has:
\begin{align*}
 (M_fs)(x) &= (\Phi^*f\Phi s)(x) \\
 &= \int_{\hat\Gamma} (f\Phi s)_\chi(x)\,d\chi \\
 &= \int_{\hat\Gamma} f(\chi)
\sum_{\gamma\in\Gamma}\chi(\gamma)\gamma_* s(\gamma^{-1}x)\,d\chi \\
 &= \sum_{\gamma\in\Gamma} \hat f(\gamma^{-1})\gamma_* s(\gamma^{-1}x)
\end{align*}
Since $f$ is a multiplication operator in each fiber it has fiber-wise
norm $\|f\|_\infty$, and so have $f$ and $M_f=\Phi^*f\Phi$.
 \end{proof}

\begin{corollary}[decomposable operators]
Conjugation by $\Phi$ defines an isomorphism between decomposable
bounded operators on $\int^\oplus_{\hat\Gamma}L^2(E_\chi)\,d\chi$ and
$\Gamma$-periodic bounded operators on $L^2(E)$.
\end{corollary}
\begin{proof} \item[]
\begin{description}
\item[``$\Rightarrow$''] A decomposable operator commutes with the
$C(\hat\Gamma)$-action, especially with $f_\gamma\in C(\hat\Gamma)$
which is defined by
\[ \hat f_\gamma(\gamma') := \begin{cases} 1,& \text{if
}\gamma=\gamma', \\ 0& \text{else.}  \end{cases} \]
By~\eqref{equ:faction} commuting with $f_\gamma$ is equivalent to
commuting with $\gamma$.
\item[``$\Leftarrow$'']  To commute with the $\Gamma$-action means to
commute with all $f_\gamma$ for $\gamma\in\Gamma$. Because of
\[ f_\gamma(\chi) = \chi(\gamma) \]
the $f_\gamma$ are just the characters $\widehat{\hat\Gamma}$ of the
compact group $\hat\Gamma$, and by the Peter-Weyl theorem (or simpler:
by the Stone-Weierstra\ss{} theorem) they are dense in $C(\hat\Gamma)$.
Since the operator norm  of $M_f$ and the  supremum norm of $f$
coincide the commutation relation follows for all $f\in C(\hat\Gamma)$
by continuity.
  \end{description}
\end{proof}

An unbounded operator is decomposable if and only if its (bounded)
resolvent is decomposable.
For a periodic symmetric elliptic operator $D$ we have a domain of
definition ${\mathcal D}(D)=C^\infty_c(X)$ on which $D$ is essentially
self-adjoint.
This domain is invariant for $D$ as well as for the $\Gamma$-action,
and  one has $[D,\gamma]=0$ for all $\gamma\in\Gamma$.
Thus all bounded functions of $D$ commute with the $\Gamma$-action, and
one has:

\begin{theorem}[decomposition of periodic operators]\label{theorem:DPO}
The closure $\bar D$ of every periodic symmetric elliptic
operator $D$ is decomposable with respect to the direct integral
of Hilbert spaces $\int^\oplus_{\hat\Gamma}L^2(E_\chi)\,d\chi$. A
core for the domain of $\bar D_\chi$ is given by
$C^\infty(E_\chi)$, and the action of  $D_\chi$ on
$C^\infty(E_\chi)\simeq C^\infty(E)^{\Gamma,\chi}$ is just the
action of $D$ as differential operator on
$C^\infty(E)^{\Gamma,\chi}$. We have $\bar
D_\chi=\overline{D_\chi}$, where
\begin{equation} D_\chi:=D|_{C^\infty(E)^{\Gamma,\chi}}
\label{equ:Dchi=D|Echi}\end{equation}
 and the closures are to be taken as operators in $L^2(E_\chi)$.
\end{theorem}
\begin{proof}
Given the remark above we have shown the decomposability already.

 $C^\infty_c(X)$ is a core for $\bar D$, its image under $\Phi_\chi$ is
contained in $C^\infty(E)^{\Gamma,\chi}$ and is a core for $\bar
D_\chi$, since $\Phi$ is an isometry.
On this domain  \eqref{equ:schi} gives the action of $\bar D_\chi$ as
asserted in the theorem.
Since $D_\chi$ is a symmetric elliptic operator on the compact manifold
$M$ it is essentially self-adjoint. $\bar D_\chi$ is a fiber of $\bar
D$ \citep[which is self-adjoint by, e.g.,][]{Ati:EODGNA} and therefore
self-adjoint, thus both define the same unique self-adjoint extension
$\overline{D_\chi}$ of $D_\chi$.
 \end{proof}

In passing we harvest a corollary which we will not use in the sequel,
but which is well known in the Euclidean setting:
\begin{corollary}[reverse Bloch property]\label{corol:RBP}
Every symmetric elliptic abelian periodic operator has the reverse
Bloch property,  i.e.\ to every $\lambda\in\spec \bar D$ there is a
bounded generalized eigensection $s\in C^\infty(E)$ with $Ds=\lambda
s.$
\end{corollary}
\begin{proof}
If $\lambda\in\spec \bar D$ then, by the general theory for direct
integrals,
\[ \{\chi\in\hat\Gamma\mid (\lambda-
\varepsilon,\lambda+\varepsilon)\cap\spec \bar D_\chi\neq\emptyset\} \]
has positive measure for every $\varepsilon>0$. The fibers $\bar
D_\chi$ are elliptic operators on a compact manifold and thus have
discrete spectrum;
the eigenvalues depend continuously on $\chi$ (even piece-wise real-
analytically; see below).
We choose a sequence $(\chi_n)_{n\in\N}$ with $(\lambda-
1/n,\lambda+1/n)\cap\spec \bar D_{\chi_n}\neq\emptyset$, so that there
is an accumulation point $\chi_\infty$ ($\hat\Gamma$ is compact), and
$\lambda\in\spec\bar D_{\chi_\infty}$ due to continuity.

Since $\spec\bar D_{\chi_\infty}$ is discrete $\lambda$ is an
eigenvalue of $\bar D_{\chi_\infty}$.
The lift of an eigensection (which is smooth due to ellipticity) lies
in $C^\infty(E)^{\Gamma,\chi}$ and therefore is bounded.
Furthermore the lift satisfies the same eigenvalue equation because of
\eqref{equ:Dchi=D|Echi}.
 \end{proof}


\section{Commutative Bloch theory from a non-commutative point of view}
\label{sec:CBTNPV}
By Gelfand's representation theorem every commutative \CS-algebra
$\mathcal A$ is isomorphic to $C_\infty(X)$,
the continuous functions vanishing at infinity of a  topological
Hausdorff space $X$,
 where  $X$ is the spectrum $\widehat{\mathcal A}$ of $\mathcal A$,
i.e.\
the set of equivalence classes of irreducible unitary representations
\cite[see
 e.g.\ ][]{Mur:CAOT}; the \CS-norm is given by the supremum norm, the
involution by point-wise complex conjugation.
Hilbert $\mathcal A$-modules are given by the sections
$C_\infty({\mathcal H})$ of a continuous field of Hilbert spaces over
$X$, finitely generated projective $\mathcal A$-modules are given by
the sections $C_\infty(E)$ of a vector bundle $E$ over $X$
\cite[]{Swa:VBPM}.
In this section we describe the corresponding structures in the case of
periodic elliptic differential operators, so that we can find a
formulation of Bloch theory that avoids using the points of the space
$\hat\Gamma$ and relies solely on the algebraic structures with respect
to $C(\hat\Gamma)$.

In Proposition~\ref{proposition:faction} we already determined the action of
$C(\hat\Gamma)$ on
$L^2(E)$. Now we use the  scalar product that is given in each fiber by
the direct integral to define a  $C(\hat\Gamma)$-values scalar product:

\begin{defiprop}[pre-Hilbert $C(\protect\hat{\Gamma})$-
module]\label{defiprop:mit dem bloeden Label}
For $s_1,s_2\in C^\infty_c(E)$ we define by
\begin{equation}
 \<s_1|s_2>(\chi) := \<(\Phi s_1)_\chi|(\Phi s_2)_\chi>_{L^2(E_\chi)}
\end{equation}
a $C(\hat\Gamma)$-valued scalar product that makes  $C_c(E)$ into a
pre-Hilbert \CS-module over $C(\hat\Gamma)$; it is a submodule of the
$C(\hat\Gamma)$-module $L^2(E)$.
\end{defiprop}
\begin{proof}
 $C_c(E)$ is obviously a $C(\hat\Gamma)$-submodule of $L^2(E)$.
Furthermore, by definition the scalar product is
{\allowdisplaybreaks
\begin{equation}
\begin{split}
 \<s_1|s_2>(\chi) &= \<(\Phi s_1)_\chi|(\Phi s_2)_\chi>_{L^2(E_\chi)}
\\
 &= \sum_{\gamma,\gamma'\in\Gamma} \bar\chi(\gamma)\chi(\gamma')
\int_{\mathcal
 D} \<\gamma_*s_1(\gamma^{-1}x)|\gamma'_*s_2(\gamma'{}^{-
1}x)>_{E_x}\,dx  \\
 &= \sum_{\gamma,\gamma'\in\Gamma} \chi(\gamma^{-1}\gamma')
\int_{\gamma^{-1}{\mathcal
 D}} \<s_1(y)|\gamma^{-1}_*\gamma'_*s_2(\gamma'{}^{-1}\gamma
y)>_{E_y}\,dy  \\
 &= \sum_{\gamma,\gamma''\in\Gamma} \chi(\gamma'') \int_{\gamma^{-
1}{\mathcal
 D}} \<s_1(y)|\gamma''_*s_2(\gamma''{}^{-1} y)>_{E_y}\,dy  \\
 &= \sum_{\gamma''\in\Gamma} \chi(\gamma'') \int_X
\<s_1(y)|\gamma''_*s_2(\gamma''{}^{-1} y)>_{E_y}\,dy  \\
 &= \sum_{\gamma''\in\Gamma} \chi(\gamma'')
\<s_1|T_{\gamma''}s_2>_{L^2(E)}
\end{split}
 \label{equ:C(hatGamma) scalar product}
\end{equation}
}
and therefore continuous  in $\chi$, since the last sum in
\eqref{equ:C(hatGamma) scalar product} is finite.
The *-property is immediately clear,  the $C(\hat\Gamma)$-linearity of
the scalar product follows from
\begin{align*}
\<s_1|M_fs_2>(\chi) &= \<(\Phi s_1)_\chi|(\Phi \Phi^*f\Phi
s_2)_\chi>_{L^2(E_\chi)} \\
&= \<(\Phi s_1)_\chi|(f\Phi s_2)_\chi>_{L^2(E_\chi)} \\
&= \<(\Phi s_1)_\chi|f(\chi)(\Phi s_2)_\chi>_{L^2(E_\chi)} \\
&= \<(\Phi s_1)_\chi|(\Phi s_2)_\chi>_{L^2(E_\chi)} f(\chi) \\
&= \<s_1|M_fs_2>(\chi) f(\chi).
\end{align*}
\end{proof}
\eqref{equ:C(hatGamma) scalar product} is the Fourier transform of the map
 $\gamma\mapsto \<s_1|T_\gamma s_2>$ and will lead us on the right
track
for the construction of a suitable Hilbert \CS-module in the non-
commutative example of gauge-periodic elliptic operators
 (see Lemma~\ref{lemma:left pre-Hilbert A-module}).

In appendix \ref{sec:HCM} we describe how -- for arbitrary (i.e.\ non-
commutative) \CS-algebras -- a \CS-valued scalar product on a $\cal A$-
module together with the \CS-norm on $\cal A$ defines a Banach norm on
the $\cal A$-module. The \CS-norm on $C(\hat\Gamma)$ is the supremum
norm, so that in this case the Banach norm  $\|\cdot\|_{\cal E}'$ on
${\cal E}':=C_c(E)\ni s$ is given by
\[ \|s\|_{{\cal E}'} := \sup_{\chi\in\hat\Gamma} \<s|s>(\chi). \]
We can take the closure ${\cal E}'$ with respect to this norm, and
hence make ${\cal E}'$ into a  \CS-module over $\hat\Gamma$:

\begin{defiprop}[Hilbert $C(\protect\hat{\Gamma})$-module and GNS
representation] \label{defiprop:HCMGR}
\begin{sloppypar} We denote the closure of $C_c(E)$ as Hilbert $C(\hat\Gamma)$-module by
${\mathcal
E}$. ${\mathcal E}$ is a submodule of the  $C(\hat\Gamma)$-module
$L^2(E)$. The Haar measure on $\hat\Gamma$ defines a faithful trace
$\tau$ on
$C(\hat\Gamma)$, and the corresponding GNS representation  $\pi_{\tau}$
(see appendix \ref{sec:HCM}) of ${\mathcal E}$ is just
the original  $C(\hat\Gamma)$-action on $L^2(E)$.

\end{sloppypar}
\end{defiprop}
\begin{proof}
Since
\[ \|\<s_1|s_2>_{\mathcal E}\|_{L^\infty(\hat\Gamma)} \geq
\|\<s_1|s_2>_{\mathcal E}\|_{L^1(\hat\Gamma)} \geq
|\<s_1|s_2>_{L^2(E)}|, \]
the closure of $C_c(E)$ in the ${\mathcal E}$-norm is a subspace of
$L^2(E)$, and by definition a $C(\hat\Gamma)$-module.

The integral with respect to a measure defines a trace. Since
$\hat\Gamma$ is compact ($\Gamma$ is discrete) it has finite volume
with respect to Haar measure, so that the trace is finite,
and all $f\in C(\hat\Gamma)\subset
L^1(\hat\Gamma)$ are trace class. Since $\hat\Gamma$ has no open
subsets of Haar measure zero the trace is faithful.
We can compute the scalar product that is defined by $\tau$ for
 $s_1,s_2\in{\mathcal E}$ as follows:
\begin{eqnarray*}
\< s_1|s_2>_{\tau} &{\overset{\text{def}}{=}}& \tau \<
s_1|s_2>_{\mathcal E} \\
 &{\overset{\text{\eqref{equ:C(hatGamma) scalar product}}}{=}}&
\int_{\hat\Gamma} \sum_{\gamma\in\Gamma} \chi(\gamma)
\<s_1|T_{\gamma}s_2>_{L^2(E)}\,d\chi \\
 &{\overset{\text{\eqref{equ:Charakter}}}{=}}& \<s_1|s_2>_{L^2(E)}
\end{eqnarray*}
Since ${\mathcal E}\supset C_c(E)$ is dense in $L^2(E)$ with respect to
the $L^2$-norm and
therefore with respect to the norm generated by  $\tau$, the GNS
representation space
for $\tau$ is $L^2(E)$. Hence, the module structures coincide.
\end{proof}

\begin{proposition}[continuous field of Hilbert spaces over
$\protect\hat{\Gamma}$]
The continuous field of Hilbert spaces over $\hat\Gamma$ that
corresponds to  ${\mathcal E}$ (see appendix  \ref{sec:CFHS}) has the
fiber $L^2(E_\chi)$ over $\chi$, the continuity structure is defined by
${\mathcal E}$.
\end{proposition}
\begin{proof}
We get the fiber at  $\chi$ as  GNS representation space of the state
$\pi_\chi:C(\hat\Gamma)\ni f\mapsto f(\chi)$. For the continuity
structure see \ref{sec:CFHS}.
\end{proof}

To sum up: We have replaced the decomposition of  $L^2(E)$ into a
direct integral of Hilbert spaces over the space
 $\hat\Gamma$ by a
Hilbert \CS-module over the \CS-algebra $C(\hat\Gamma)$, endowed with a
faithful trace whose GNS representation gives us back the original
Hilbert space $L^2(E)$.
In Proposition~\ref{proposition:faction} we determined the $C(\hat\Gamma)$-action
and noticed that decomposable bounded operators with respect to the
direct integral are just the ones commuting with this action (the
periodic operators).
This, decomposable operators are just the module maps on the
$C(\hat\Gamma)$-module
$L^2(E)$. This includes especially the images (under the GNS
representation) of module maps on ${\mathcal E}$. To conclude this
section we cite a special case of Theorem~\ref{theorem:GPMO} from
Section~\ref{sec:A} that shows that periodic elliptic differential
operators define indeed regular unbounded  module maps \cite[see
e.g.\
][chapter~9 for these notions]{Lan:HCMTOA} on ${\mathcal E}$, so that
the resolvent of such operators belongs to the image of the
 GNS representation.

\begin{theorem}[decomposition of periodic operators]
Let  $D$ be a periodic symmetric elliptic differential operator.
Then $D$ defines a regular operator $D_{\mathcal E}$ with domain
of definition ${\mathcal D}(D_{\mathcal E})=C^\infty_c(E)$ in
${\mathcal E}$. For $\lambda\in\R$ we have
\begin{equation} \pi_{\tau}\left( (\lambda\ONE_{\mathcal
E}+\overline{D_{\mathcal E}})^{-1}\right) =
(\lambda\ONE_{L^2(E)}+\bar D)^{-1}. \end{equation}
\end{theorem}


\section{Non-commutative Bloch theory} \label{sec:NBT}

Motivated by the non-commutative insight gained in the previous
section, we will now define a general class of abstract elliptic
operators that allows for a non-commutative version of Bloch theory.
This will let us read off spectral properties from properties of the
\CS-algebras that are involved.

\begin{definition}[${\cal A}$-elliptic operator]
\label{definition:A-elliptic operator}
Let ${\mathcal A}$ be a  unital \CS-algebra, ${\mathcal E}$ a Hilbert
\CS-module over ${\mathcal A}$.
An unbounded operator $D$ on ${\mathcal E}$ is called \emph{${\mathcal
A}$-elliptic} if
\begin{enumerate}
\item $D$ is densely defined,
\item $D$ is regular in the sense that $D$ has a densely defined
adjoint $D^*$ with range $\ran(1+D^*D)\overset{\text{dense}}{\subset}{\cal
E}$, \\\hglue-\leftmargin and
\item $D$ has ${\cal A}$-compact resolvent\footnote{
We will explain this name in the proof of Lemma \ref{lemma:spectral
projections}.},   i.e.\ $(1+D^*D)^{-1}\in {\mathcal K}_{\mathcal
A}({\mathcal E})$.
\end{enumerate}
\end{definition}

Hilbert modules are understood to be Hilbert right modules, as
described in appendix \ref{sec:HCM}.
Hilbert spaces are Hilbert  $\C$-modules, therefore our scalar products
are complex linear in the second entry and complex anti-linear in the
first entry, corresponding to the convention in Mathematical Physics.

\begin{remark}[module und Hilbert space operators]
Given a normalized faithful trace $\tau$ on ${\mathcal A}$ we can define,
as described in appendix \ref{sec:HCM}, a Hilbert space scalar product
on ${\cal E}$ by
\[ \<e_1|e_2>_\tau := \tau(\<e_1|e_2>_{\cal E}) \]
for $e_1,e_2\in {\cal E}$.
Let $H_\tau$ be the completion of ${\cal E}$ with respect to
$\<\cdot|\cdot>_\tau$,  i.e.\ the corresponding GNS representation
space.
We write $\<\cdot|\cdot>_{H_\tau}$ for $\<\cdot|\cdot>_\tau$.
${\cal L}_{\cal A}({\cal E})$ is represented faithfully on  $H_\tau$.
Thus the spectrum of an element  $a$ of the  \CS-algebra ${\cal
L}_{\cal A}({\cal E})$ coincides (as a set) with the spectrum of the
operator $\pi_\tau(a)$ on the Hilbert space $H_\tau$:
\end{remark}
\begin{lemma}[spectrum of module and Hilbert space operators]
If $a\in{\cal L}_{\cal A}({\cal E})$ then
\[ \spec a = \spec \pi_\tau(a). \]
\end{lemma}
In the sequel we will identify  ${\mathcal E}$ resp.\ ${\cal L}_{\cal
A}({\cal E})$ with the images in $H_\tau$ resp.\ ${\cal L}(H_\tau)$.

\begin{defiprop}[$\tr_\tau$-trace]\label{defiprop:tr_tau-trace}
On the  $\cal A$-finite operators\footnote{
 \begin{align*}{\cal F}_{\cal A}({\cal
E})&=\operatorname{span}\{\pi^{\cal E}_{x,y}\mid x,y\in{\cal E}\}\text{
with} \\
\pi^{\cal E}_{x,y}(z) &= x\<y|z>_{\cal E}\text{ for }z\in{\cal E},
\end{align*}
so that ${\cal K}_{\cal A}({\cal E})=\overline{{\cal F}_{\cal A}({\cal
E})}$.}
 ${\cal F}_{\cal A}({\cal E})$ we define a faithful trace by
\begin{equation} \tr_{\tau}(\pi^{\cal E}_{x,y}) =
\tau\left(\<y|x>_{\cal E}\right), \end{equation}
the trace associated to  $\tau$ in the GNS representation.
We denote the corresponding trace class ideal  in ${\cal L}_{\cal
A}({\cal E})$ by ${\cal L}^1_{\cal A}({\cal E},\tr_\tau)$.
\end{defiprop}
\begin{proof}
For the generators of  ${\cal F}_{\cal A}({\cal E})$ one can easily
show the relations
\begin{alignat*}{2}
\left(\pi^{\cal E}_{x,y}\right)^* &= \pi^{\cal E}_{y,x}, & \quad
\pi^{\cal E}_{xa,y} &= \pi^{\cal E}_{x,ya^*}, \\
T\pi^{\cal E}_{x,y} &= \pi^{\cal E}_{Tx,y}, & \quad
\pi^{\cal E}_{x,y}T &= \pi^{\cal E}_{x,T^*y}
\end{alignat*}
for $x,y\in{\cal E},a\in{\cal A},T\in{\cal L}_{\cal A}({\cal E})$.
Thus, from the trace property of $\tau$ we have
\begin{align*}
\tr_{\tau}\left((\pi^{\cal E}_{x,y})^*\right) &= \tau\left(\<x|y>_{\cal
E} \right)\\
 &= (\tr_{\tau}\pi^{\cal E}_{x,y})^*, \\
\tr_{\tau}(T\pi^{\cal E}_{x,y}) &= \tau\left(\<y|Tx>_{\cal E}\right) \\
 &= \tau\left(\<T^*y|x>_{\cal E}\right) \\
 &= \tr_{\tau}(\pi^{\cal E}_{x,y}T).
\end{align*}
For all $z,t\in{\cal E}$ we have
\begin{align*} \tr_{\tau}(\pi^{\cal E}_{x,y}\pi^{\cal E}_{z,t})
&=\tr_{\tau}\left(\pi^{\cal E}_{x\<y|z>_{\cal E},t}\right) \\
 &= \tau\left(\<t|{x\<y|z>_{\cal E}}>_{\cal E}\right) \\
 &= \tau\left(\<t|\pi^{\cal E}_{x,y}(z)>_{\cal E}\right)
\end{align*}
so that $\tr_{\cal E}$ is faithful: Set $t=\pi^{\cal E}_{x,y}(z)$, and note
that $\tau$ is a faithful trace on $\cal A$.
\end{proof}

\begin{remark}[$\tr\pi^{ H_\tau}_{x,y}$ versus $\tr_\tau\pi^{\cal
E}_{x,y}$]
By Definition \ref{defiprop:tr_tau-trace} we have
\[\tr_\tau \pi^{\cal E}_{x,y}=\tau\left(\<y|x>_{\cal E} \right) =
\<y|x>_{H_\tau}=\tr\pi^{ H_\tau}_{x,y}\]
with the usual canonical Hilbert space trace $\tr$ and the usual rank 1
operators
\begin{align*} \pi^{ H_\tau}_{x,y}:H_\tau\ni z&\mapsto
x\<y|z>_{H_\tau}\in H_\tau
\end{align*}
on the Hilbert space $H_\tau$. However,  $\pi^{\cal E}_{x,y}$ and
$\pi^{ H_\tau}_{x,y}$ are different  operators:
\begin{align*}
\pi^{\cal E}_{x,y}(z) &= x\<y|z>_{\cal E}\\
\intertext{whereas}
\pi^{H_\tau}_{x,y}(z) &= x\<y|z>_{H_\tau} \\
 &= x\tau\left(\<y|z>_{\cal E}\right).
\end{align*}
Thus, in general  $\tr_\tau$ and $\tr$ are indeed different traces.
\end{remark}

\begin{remark}[$\tr\pi^{\cal E}_{x,y}$ versus $\tr_\tau\pi^{\cal
E}_{x,y}$]
Let $(e_n)_{n\in\N}$ be an orthonormal base\footnote{To simplify matters
we assume  $e_n\in{\cal E}$ for all $n\in\N$. Since $\cal E$ is dense
in $H_\tau$ this can always be achieved.} of $H_\tau$. Then
{\allowdisplaybreaks \begin{align*}
\tr_\tau\pi^{\cal E}_{x,y} &= \tr \pi^{ H_\tau}_{x,y} \\
 &= \sum_{n\in\N}\<e_n|\pi^{H_\tau}_{x,y}(e_n)>_{H_\tau} \\
 &= \sum_{n\in\N}\<e_n|x{\<y|e_n>_{H_\tau}}>_{H_\tau} \\
 &= \sum_{n\in\N}\<e_n|x>_{H_\tau}\<y|e_n>_{H_\tau} \\
 &= \sum_{n\in\N}\tau\left(\<e_n|x>_{\cal
E}\right)\tau\left(\<y|e_n>_{\cal E}\right), \\
\tr\pi^{\cal E}_{x,y} &=\sum_{n\in\N}\<e_n|\pi^{\cal
E}_{x,y}(e_n)>_{H_\tau} \\
 &= \sum_{n\in\N}\<e_n|x{\<y|e_n>_{\cal E}}>_{H_\tau} \\
 &= \sum_{n\in\N}\tau\left( \<e_n|x{\<y|e_n>_{\cal E}}>_{\cal E}
\right) \\
 &= \sum_{n\in\N}\tau\left( \<e_n|x>_{\cal E}\<y|e_n>_{\cal E}
\right).
\end{align*}}
So, $\tr \pi^{ H_\tau}_{x,y}$ and  $\tr\pi^{\cal E}_{x,y}$ coincide if
$\tau$ is multiplicative.
But in this case $\tau$, being a multiplicative faithful trace, is a *-
isomorphism ${\cal A}\rightarrow \C$ already, so that we just reproduce
the Hilbert space trace.

In general  $\tr$ will be larger than $\tr_\tau$ because
\begin{align*}
\tr\pi^{\cal E}_{e_m,e_m} &= \sum_{n\in\N}\tau\left( \<e_n|e_m>_{\cal
E}\<e_m|e_n>_{\cal E} \right)\\
  &= \sum_{n\in\N}\tau\left( \<e_m|e_n>_{\cal E}^*\<e_m|e_n>_{\cal E}
\right)\\
&\geq \tau\left( \<e_m|e_m>_{\cal E}^*\<e_m|e_m>_{\cal E} \right)\\
&\geq \left(\tau(\<e_m|e_m>_{\cal E}) \right)^2 \\
&= \|e_m\|^2_{H_\tau} \\
&= 1\\
&= \tr\pi_{e_m,e_m}^{H_\tau}\\
&= \tr_\tau\pi^{\cal E}_{e_m,e_m}.
\end{align*}
Here we used the Cauchy-Schwarz inequality
$\tau(a^*b)\leq\sqrt{\tau(a^*a)\tau(b^*b)}$ and the fact that the trace
is normalized.

To sum up: The $\tr_\tau$-trace is defined only on the image of the
adjointable module operators in the GNS representation, and on these it
is in general smaller than the Hilbert space trace, so that the
corresponding trace class ideal is larger:
\begin{align*}
 \pi_\tau\left({\cal L}^1_{\cal A}({\cal E},\tr_\tau)\right) \supset
\pi_\tau\left({\cal L}_{\cal A}({\cal E})\right)\cap {\cal
L}^1(H_\tau,\tr)
\end{align*}
\end{remark}

\begin{remark}[$\tr_\tau$ for standard Hilbert modules]
If $\cal E$ is a standard $\cal A$-module $H\otimes{\cal A}$ (tensor
product of Hilbert modules) with a Hilbert space $H$ then the  GNS representation
space  $H_\tau$ of $\cal E$ is given by $H_\tau=H\otimes h_\tau$
(tensor product of Hilbert spaces), where $h_\tau$ is the GNS
representation space of $\cal A$.
Therefore we have for the elementary tensors $x\otimes a,y\otimes
b\in{\cal E}$
\begin{align*}
 \<y\otimes b|x\otimes a>_{\cal E}
 = \<y|x>_H \,b^*a, \\
\pi^{H_\tau}_{x\otimes a,y\otimes b} &=
\pi^{H}_{x,y}\otimes\pi^{h_\tau}_{a,b},\\
\pi^{\cal E}_{x\otimes a,y\otimes b} &= \pi^{H}_{x,y}\otimes\pi^{\cal
A}_{a,b} \\
 &= \pi^{H}_{x,y}\otimes  ab^*. \\
\intertext{With the standard traces $\tr_H,\tr_{h_\tau}$ on the Hilbert
spaces $H,h_\tau$ we get}
\tr \pi^{H_\tau}_{x\otimes a,y\otimes b} &= \tr_{\tau}\pi^{\cal
E}_{x\otimes a,y\otimes b} \\
 &= \<y|x>_H \,\tau(b^*a) \\
 &= \tr_H \left(\pi^{H}_{x,y}\right)\, \tr_{h_\tau}
\left(\pi^{h_\tau}_{a,b}\right).  \\
\intertext{Thus we arrive at}
 \tr &= \tr_H\otimes \tr_{h_\tau}, \\
 \tr_\tau &= \tr_H\otimes \operatorname{\tau}.
\end{align*}
\end{remark}

\begin{lemma}[$\tr$ for $\tr_\tau$-trace class]\label{lemma:tr for
trtau-trace class}
Let  ${\cal E}=H\otimes{\cal A}$ be as above.
If $\cal A$ is infinite dimensional with a unitary orthonormal basis
for $h_\tau$,
then  $0$ is the only  $\tr_\tau$-trace class operator with finite
standard trace. In particular: All Hilbert $\cal A$-submodules are
infinite dimensional vector spaces.
\end{lemma}
\begin{proof}
Let $(x_n)_{n\in\N}$ be an orthonormal basis of $h_\tau$, consisting of
unitary elements of $\cal A$.
Then
\begin{align*}
\tr_{h_\tau}\pi^{\cal A}_{a,b} &= \sum_{n\in\N} \<x_n|ab^*x_n>_{h_\tau}
\\
&= \sum_{n\in\N} \tau(x_n^*ab^*x_n) \\
&= \sum_{n\in\N} \tau(ab^*).
\end{align*}
\end{proof}

\begin{lemma}[non-existence of finite dimensional
modules]\label{lemma:non-existence of finite dimensional modules}
If $\cal A$ is infinite dimensional with a unitary orthonormal basis for
$h_\tau$ then every projective $\cal A$-module is an infinite
dimensional vector space.
\end{lemma}
\begin{proof}
If $\cal E$ is a projective Hilbert $\cal A$-module then $\cal E$ is a
direct summand of a free module $H\otimes{\cal A}$ for a suitable
Hilbert space $H$, and we can apply Lemma \ref{lemma:tr for trtau-trace
class}.
\end{proof}


\begin{lemma}[spectral projections]\label{lemma:spectral projections}
Let  $D$ be a self-adjoint
$\cal A$-elliptic operator and  let $\lambda_1,\lambda_2\in\R\setminus\spec
D$, $\lambda_1\leq\lambda_2$.
Then the corresponding spectral projection\footnote{If
$\lambda_1,\lambda_2\in\R\setminus\spec D$ then
$P_{[\lambda_1,\lambda_2]}=P_{(\lambda_1,\lambda_2]}=P_{[\lambda_1,
\lambda_2)}$.} $P_{[\lambda_1,\lambda_2]}$ on the interval
$[\lambda_1,\lambda_2]$ is $\cal A$-compact.
If $e^{-tD^2}\in{\cal L}^1_{\cal A}({\cal E},\tr_\tau)$ for $t>0$ then
the spectral projections are $\tr_\tau$-trace class.
\end{lemma}
\begin{proof} \item[]
\begin{description}
\item[Reduction to $D\geq0$]
If $\spec D=\R$ there is nothing to prove. So, let
$\lambda_0\in\R\setminus\spec D$.
We show that we can assume $D\geq0$ for the proof of $\cal A$-
compactness: Let
\begin{align*} D'&:= f(D)\text{ with}\\ f(x)&:=x-\lambda_0\text{ for
}x\in\R. \end{align*}
Then $0\notin\spec D'$. We set $g(x):=(1+x^2)^{-1}$ so that
\begin{align*}
(1+D'{}^2)^{-1} &= g\circ f(D)\\
 &= g(D)b(D)\text{ with}\\
b(x)&=\frac{g\circ f(x)}{g(x)}\\
 &=\frac{1+(x-\lambda_0)^2}{1+x^2}.
\end{align*}
Since  $b$ is continuous and bounded $b(D)\in{\cal L}_{\cal A}({\cal
E})$.
If  $D$ is $\cal A$-elliptic, i.e.\ $g(D)\in{\cal K}_{\cal A}({\cal
E})$, then we get  $g(D)b(D)\in{\cal K}_{\cal A}({\cal E})$,   i.e.\
$D'$ is $\cal A$-elliptic.
Denote the spectral projections of $D'$ with $P'$.
Then obviously $P'(\lambda)=P(\lambda+\lambda_0)$, so that it suffices
to test  $P'$ for $\cal A$-compactness.

Finally we set $D'':=|D'|$.
Then  $D''$ is $\cal A$-elliptic by definition, positive by
construction, and strictly positive because $0\notin\spec D'$.
If we denote the spectral projections of $D''$ by $P''$ then
\begin{align*}
P''(\lambda) &= 1_{(-\infty,\lambda]}(D'') \\
 &=\left(1_{(-\infty,\lambda]}\circ|\cdot|\right)(D') \\
 &=1_{[-\lambda,\lambda]}(D')\\
 &=P'_{[-\lambda,\lambda]}.\\
\intertext{Therefore we get for $0\leq\lambda_1\leq\lambda_2$}
P''_{(\lambda_1,\lambda_2]} &=P''(\lambda_2)-P''(\lambda_1)\\
 &=P'_{[-\lambda_2,\lambda_2]}-P'_{[-\lambda_1,\lambda_1]}\\
 &=P'_{[-\lambda_2,-\lambda_1)\cup(\lambda_1,\lambda_2]}
\end{align*}
By assumption $0\notin\spec D'$ and therefore $P'_{[0,\infty)},P'_{(-
\infty,0]}\in{\cal L}_{\cal A}({\cal E})$, so that
\begin{align}
P'_{(\lambda_1,\lambda_2]}&=P''_{(\lambda_1,\lambda_2]}P'_{[0,\infty)}\
in{\cal K}_{\cal A}({\cal E}) \text{ and} \label{equ:P'P''+}\\
P'_{[-\lambda_2,-\lambda_1)}&=P''_{(\lambda_1,\lambda_2]}P'_{(-
\infty,0]}\in{\cal K}_{\cal A}({\cal E}), \label{equ:P'P''-}
\end{align}
if $P''_{(\lambda_1,\lambda_2]}\in{\cal K}_{\cal A}({\cal E})$. If
$\lambda_1\leq0\leq\lambda_2$ we write
\[ P'_{[\lambda_1,\lambda_2]}=P'_{[\lambda_1,0)}+P'_{(0,\lambda_2]} \]
and apply equation \eqref{equ:P'P''+} and \eqref{equ:P'P''-}. Hence it
suffices to test $P''$ for $\cal A$-compactness.

\item[$\cal A$-compactness]
We show that every spectral projection $P_{[\lambda_1,\lambda_2]}$ for
$\lambda_1,\lambda_2\in\R\setminus\spec D$ can be produce by continuous
functional calculus from  $S:=(1+D^2)^{-1}$, so that it belongs to
${\mathcal K}_{\mathcal A}({\mathcal E})$.
For this we note that  $S^{-1}=D^2+1$ is densely defined ($D$ is
regular), self-adjoint and bounded below by $1$.
Thus  $\sqrt{S^{-1}-1}$ exists, is positive and self-adjoint.
By the spectral mapping theorem we have
\begin{align*}
 z\in\spec\sqrt{S^{-1}-1}&\Leftrightarrow (z^2+1)^{-1}\in\spec S \\
 &\Leftrightarrow z\in\spec D.
\end{align*}
Therefore, the operator
 \[ R_z:=\left(z- \sqrt{S^{-1}-1}\right)^{-1}.\]
exists for all $z$ in the resolvent set of $D$.
Since the function
\[ \lambda\mapsto \left(z- \sqrt{\lambda^{-1}-1}\right)^{-1} \]
is continuous and bounded on every closed set not containing
$(z^2+1)^{-1}$,  $R_z$ belongs to the \CS-algebra generated by $S$ for
every $z\in\C\setminus\spec D$ and therefore belongs to ${\mathcal
K}_{\mathcal A}({\mathcal E})$, i.e.\ it is $\cal A$-compact.
Since
\[ P_{[\lambda_1,\lambda_2]} = \frac1{2\pi\imath}\oint_c R_z\,dz \]
for a suitable closed path $c$  in $\C\setminus\spec D$ with winding
number $1$ fulfilling $c\cap\R=\{\lambda_1,\lambda_2\}$,
$P_{[\lambda_1,\lambda_2]}$ belongs to the \CS-algebra generated by all
$R_z$.

\item[trace class property] Let  $e^{-tD^2}$ be $\tr_\tau$-trace
class\footnote{We don't assume positivity of $D$ any more.}.
Since
\begin{align*}
P_{[\lambda_1,\lambda_2]} &= \int_{\lambda_1}^{\lambda_2} dP(\lambda)
\\
 &\leq e^{t(\lambda_2-\lambda_1)^2} \int_{\lambda_1}^{\lambda_2} e^{-
t(\lambda-\lambda_1)^2} dP(\lambda) \\
 &\leq e^{t(\lambda_2-\lambda_1)^2} \int_\R e^{-t(\lambda-\lambda_1)^2}
dP(\lambda) \\
 &= e^{t(\lambda_2-\lambda_1)^2} e^{-tD^*D}
\end{align*}
the spectral projections inherit the trace class property from $e^{-
tD^2}$.
\end{description}
\end{proof}

If  $\lambda$ is an isolated eigenvalue then for sufficiently small
$\varepsilon>0$ $P_\lambda:=P_{[\lambda-
\varepsilon,\lambda+\varepsilon]}$
is the projection on the eigenspace of $\lambda$, independent of
$\varepsilon$.
So $P_\lambda$ fulfills the hypotheses of Lemma
\ref{lemma:spectral projections}, and we can determine the dimension
of the eigenspace:

\begin{theorem}[isolated eigenvalue]\label{theorem:isolated eigenvalue}
If $\lambda$ is an isolated eigenvalue of a self-adjoint $\cal A$-
elliptic operator $D$ then the corresponding eigenspace $E_\lambda$ is
an (algebraically) finitely generated projective Hilbert $\cal A$-
module, and the projection $P_\lambda$ is $\cal A$-finite.
If $e^{-tD^2}$ is $\tr_\tau$-trace class then so is $P_\lambda$,
i.e.\ $E_\lambda$ has finite $\tau$-dimension $\tr_\tau P_\lambda$.

If $\cal E,A$ fulfill the hypotheses of Lemma~\ref{lemma:non-existence
of finite dimensional modules} then $E_\lambda$ has infinite Hilbert
dimension $\tr P_\lambda$ for every isolated eigenvalue $\lambda$ of
$D$. In particular:  $D$ has essential spectrum only.
\end{theorem}
\begin{proof}
$P_\lambda$ is the spectral projection of a self-adjoint operator and
therefore self-adjoint, and $\cal A$-compact by
Lemma~\ref{lemma:spectral projections}.
Thus the eigenspace  $E_\lambda$ is the image of a closed adjointable
projection $P_\lambda$ and therefore a closed complementable  $\cal A$-
module.
Since the projection $P_\lambda|_{E_\lambda}=\ONE$ is $\cal A$-compact $E_\lambda$ is
algebraically finitely generated and projective, because algebraically
finitely generated $\cal A$-modules $E$ are just the ones with unital
${\cal K}_{\cal A}(E)$ and automatically projective \cite[see e.g.\
][Theorem 15.4.2 and Corollary 15.4.8]{Weg:KTCAFA}.

If $e^{-tD^2}\in{\cal L}_{\cal A}^1({\cal E},\tr_\tau)$ then so is
$P_\lambda$ by Lemma~\ref{lemma:spectral projections}, and under the
same hypotheses we can apply Lemma~\ref{lemma:non-existence of finite
dimensional modules}.
\end{proof}

The main idea of the following proof goes back to \cite{Sun:GCASPSOM}:

\begin{theorem}[band structure]\label{theorem:band structure}
Assume that ${\mathcal K}_{\mathcal A}({\mathcal E})$ has the Kadison
property with respect to  $\tr_\tau$ (see Definition~\ref{definition:Kadison
property}). Then the spectrum of every self-adjoint $\cal A$-elliptic
operator $D$ with $e^{-tD^2}\in{\cal L}^1_{\cal A}({\cal E},\tr_\tau)$
has band structure.
\end{theorem}
\begin{proof}
Let $a=\lambda_0<\ldots<\lambda_n=b\in\R\setminus\spec D$, so that
$P_{[\lambda_i,\lambda_{i+1}]}\neq0$ for $0\leq i\leq n-1$, i.e.\
$\spec D$ has at least  $n$ components in $[a,b]$. Then
\begin{align*}
P_{[a,b]}&= \sum_{i=0}^{n-1}P_{[\lambda_i,\lambda_{i+1}]} \\
\intertext{and therefore}
\tr_\tau P_{[a,b]}&= \sum_{i=0}^{n-1}\tr_\tau
P_{[\lambda_i,\lambda_{i+1}]} \\
 &\geq n c_K. \\
\Leftrightarrow n&\leq \frac1{c_K} \tr_\tau P_{[a,b]},
\end{align*}
since all projections occurring in this sum are  $\tr_\tau$-trace class
by Lemma~\ref{lemma:spectral projections}.
\end{proof}

If  $c_K=0$ then we cannot apply Theorem~\ref{theorem:band structure}.
Instead, spectra with the structure of a Cantor set
seem possible.
Examples show that the opening of gaps which are allowed depends
heavily on the specific structure of the operator and cannot easily be
controlled globally.
To get generic results we therefore have to make sure that not only
$c_K=0$ , but also that the trace can be arbitrarily small on ``many''
projections. This is accomplished by the following theorem by
\cite{ChoEll:DSAEFSIRCA}:

\begin{theorem}[Cantor spectrum]\label{theorem:Cantor spectrum}
Let $\cal A$ be a \CS-algebra with a faithful state $\Phi$.
Assume that every self-adjoint element can be approximated arbitrarily
well by an element with finite spectrum on whose minimal spectral
projections  $\Phi$ is arbitrarily small.
Then the self-adjoint elements with Cantor spectrum are dense in all
self-adjoint elements.
\end{theorem}

In particular, the algebras in Theorem~\ref{theorem:Cantor spectrum}
have real rank zero, i.e.\ the invertible self-adjoint elements are
dense in all self-adjoint ones:
\begin{defiprop}[real rank]\label{defiprop:real rank}
Let $\cal A$ be a unital  \CS-algebra. The \emph{real rank of
$\cal A$} is defined by
\begin{multline*}
\begin{aligned}
\RR({\cal A}) &=\min\{m\in\N_0\mid \forall{n\geq m+1}:\RR_n({\cal
A})\}, \text{ where} \\
\RR_n({\cal A}) &=  \Bigg\langle \forall x\in{\cal
A}_{sa}^n:\forall{\varepsilon>0}:\exists y\in{\cal A}_{sa}^n:
\end{aligned} \\
\sum_{k=1}^ny_k^2\in{\cal A}^\times\wedge \Bigg\|\sum_{k=1}^n(y_k-
x_k)^2\Bigg\|<\varepsilon\Bigg\rangle
\end{multline*}
For all $n\in\N_0$ we have $\RR_n({\cal A})\Rightarrow\RR_{n+1}({\cal
A})$.
The following conditions are equivalent:
\begin{enumerate}
\item $\RR({\cal A})=0$ \label{enum:RRA=0}
\item ${\cal A}_{sa}^\times\subset{\cal A}_{sa}$ dense
\label{enum:Asaxdense}
\item The self-adjoint elements with finite spectrum are dense in ${\cal
A}_{sa}$. \label{enum:AsaFSdense}
\end{enumerate}
We say  $\cal A$ \emph{has real rank 0 with infinitesimal state}
(\emph{$RRI_0$}) if  $\cal A$ fulfills the assumptions of
Theorem~\ref{theorem:Cantor spectrum}.
\end{defiprop}
For the convenience of the reader we include a proof of these
equivalences which are well known in the \CS-community.
\begin{proof}
\item[] \begin{description}
\item[$\RR_n({\cal A})\Rightarrow\RR_{n+1}({\cal A})$:] Let $x\in{\cal
A}_{sa}^{n+1}$ and $\tilde x:=(x_1,\ldots,x_n)\in{\cal A}_{sa}^n$. By
assumption there is  $\tilde y\in{\cal A}_{sa}^n$ such that
\[ \sum_{k=1}^n\tilde y_k^2\in{\cal A}^\times\wedge
\left\|\sum_{k=1}^n(\tilde y_k-x_k)^2\right\|<\varepsilon.\]
For all $k=1,\ldots,n+1$ we have $x_k^2\geq0$, and $\sum_{k=1}^n\tilde
y_k^2>0$.
 We set  $ y:=(\tilde y_1,\ldots,\tilde y_n,x_{n+1} )$.
Then
\begin{align*} \sum_{k=1}^{n+1}y_k^2 &\geq \sum_{k=1}^n y_k^2 =
\sum_{k=1}^n\tilde y_k^2 >0\\
\Rightarrow \sum_{k=1}^{n+1}y_k^2&\in{\cal A}^\times
\intertext{and finally}
\left\|\sum_{k=1}^{n+1}(y_k-x_k)^2\right\|=\left\|\sum_{k=1}^n(\tilde
y_k-x_k)^2\right\| <\varepsilon
\end{align*}
\item[\ref{enum:RRA=0} $\Rightarrow$ \ref{enum:Asaxdense}] by
definition.
\item[\ref{enum:Asaxdense} $\Rightarrow$ \ref{enum:RRA=0}] because
$\RR_0({\cal A})\Rightarrow\RR_{n+1}({\cal A})$ for all $n\in\N_0$.
\item[\ref{enum:AsaFSdense} $\Rightarrow$ \ref{enum:Asaxdense}:] Let
$x\in{\cal A}_{sa}$, $\varepsilon>0$.
Then there is $y\in{\cal A}_{sa}$ with finite spectrum so that $\|x-
y\|<\varepsilon/2$.
If  $y$ is invertible then there is nothing to prove, otherwise we
choose $\delta>0$ so that $\spec y\cap B_\delta(0)=\{0\}$.
Then $\tilde y:=y+\frac12\min\{\delta,\varepsilon\}\ONE$ is invertible,
and \[ \|\tilde y-x\|\leq\|y-x\|+\varepsilon/2<\varepsilon.\]
\item[\ref{enum:Asaxdense} $\Rightarrow$ \ref{enum:AsaFSdense}:]
Let $x\in{\cal A}_{sa}$. We show first that we can approximate $x$ by a
self-adjoint element with finitely many gaps.
For this we choose a sequence $(s_i)_{i\in\N}$ of pair-wise distinct real
numbers that are dense in the interval $[-\|x\|,\|x\|]$.
We set $x_1:=x$ and choose inductively $y_n\in{\cal A}_{sa}^\times$ such
that
\begin{align*}
\|(x_n-s_n\ONE)-y_n\| &< 2^{-
n}\min\{\varepsilon,\varepsilon_1,\ldots,\varepsilon_{n-1}\},\text{
where} \\
 \varepsilon_i&=\dist\{0,\spec y_i\}>0, \\
\intertext{since $y_i$ is invertible, and we define}
x_{n+1}:&=y_n+s_n\ONE.
\intertext{Then, by construction}
\|x_{n+1}-x_n\|&< 2^{-
n}\min\{\varepsilon,\varepsilon_1,\ldots,\varepsilon_{n-1}\} \\
\Rightarrow \|x_{n+1}-x\|&\leq \sum_{i=1}^n\|x_{i+1}-x_i\| \\
 &< \sum_{i=1}^n 2^{-
i}\min\{\varepsilon,\varepsilon_1,\ldots,\varepsilon_{i-1}\} \\
 &\leq \min\{\varepsilon,\varepsilon_1,\ldots,\varepsilon_{n-1}\}.
\end{align*}

\begin{figure}[htbp]
\[ \hbox{
\setlength{\unitlength}{0.240900pt}
\ifx\plotpoint\undefined\newsavebox{\plotpoint}\fi
\sbox{\plotpoint}{\rule[-0.200pt]{0.400pt}{0.400pt}}%
\begin{picture}(750,450)(0,0)
\font\gnuplot=ptmr7t at 10pt
\gnuplot
\sbox{\plotpoint}{\rule[-0.200pt]{0.400pt}{0.400pt}}%
\put(176.0,68.0){\rule[-0.200pt]{122.859pt}{0.400pt}}
\put(176.0,68.0){\rule[-0.200pt]{0.400pt}{86.483pt}}
\put(176.0,158.0){\rule[-0.200pt]{4.818pt}{0.400pt}}
\put(154,158){\makebox(0,0)[r]{$\frac{\tilde s_1+\tilde s_2}2$}}
\put(666.0,158.0){\rule[-0.200pt]{4.818pt}{0.400pt}}
\put(176.0,248.0){\rule[-0.200pt]{4.818pt}{0.400pt}}
\put(154,248){\makebox(0,0)[r]{$\frac{\tilde s_2+\tilde s_3}2$}}
\put(666.0,248.0){\rule[-0.200pt]{4.818pt}{0.400pt}}
\put(176.0,337.0){\rule[-0.200pt]{4.818pt}{0.400pt}}
\put(154,337){\makebox(0,0)[r]{$\frac{\tilde s_3+\tilde s_4}2$}}
\put(666.0,337.0){\rule[-0.200pt]{4.818pt}{0.400pt}}
\put(240.0,68.0){\rule[-0.200pt]{0.400pt}{4.818pt}}
\put(240,23){\makebox(0,0){$\tilde s_1$}}
\put(240.0,407.0){\rule[-0.200pt]{0.400pt}{4.818pt}}
\put(367.0,68.0){\rule[-0.200pt]{0.400pt}{4.818pt}}
\put(367,23){\makebox(0,0){$\tilde s_2$}}
\put(367.0,407.0){\rule[-0.200pt]{0.400pt}{4.818pt}}
\put(495.0,68.0){\rule[-0.200pt]{0.400pt}{4.818pt}}
\put(495,23){\makebox(0,0){$\tilde s_3$}}
\put(495.0,407.0){\rule[-0.200pt]{0.400pt}{4.818pt}}
\put(622.0,68.0){\rule[-0.200pt]{0.400pt}{4.818pt}}
\put(622,23){\makebox(0,0){$\tilde s_4$}}
\put(622.0,407.0){\rule[-0.200pt]{0.400pt}{4.818pt}}
\put(176.0,68.0){\rule[-0.200pt]{122.859pt}{0.400pt}}
\put(686.0,68.0){\rule[-0.200pt]{0.400pt}{86.483pt}}
\put(176.0,427.0){\rule[-0.200pt]{122.859pt}{0.400pt}}
\put(176.0,68.0){\rule[-0.200pt]{0.400pt}{86.483pt}}
\put(176,68){\usebox{\plotpoint}}
\multiput(214.58,68.00)(0.499,0.704){125}{\rule{0.120pt}{0.663pt}}
\multiput(213.17,68.00)(64.000,88.625){2}{\rule{0.400pt}{0.331pt}}
\put(176.0,68.0){\rule[-0.200pt]{9.154pt}{0.400pt}}
\multiput(329.58,158.00)(0.498,0.884){99}{\rule{0.120pt}{0.806pt}}
\multiput(328.17,158.00)(51.000,88.327){2}{\rule{0.400pt}{0.403pt}}
\put(278.0,158.0){\rule[-0.200pt]{12.286pt}{0.400pt}}
\multiput(482.58,248.00)(0.498,1.176){73}{\rule{0.120pt}{1.037pt}}
\multiput(481.17,248.00)(38.000,86.848){2}{\rule{0.400pt}{0.518pt}}
\put(380.0,248.0){\rule[-0.200pt]{24.572pt}{0.400pt}}
\multiput(597.58,337.00)(0.498,1.190){73}{\rule{0.120pt}{1.047pt}}
\multiput(596.17,337.00)(38.000,87.826){2}{\rule{0.400pt}{0.524pt}}
\put(520.0,337.0){\rule[-0.200pt]{18.549pt}{0.400pt}}
\put(635.0,427.0){\rule[-0.200pt]{12.286pt}{0.400pt}}
\put(176,68){\usebox{\plotpoint}}
\put(176.00,68.00){\usebox{\plotpoint}}
\multiput(181,72)(17.798,10.679){0}{\usebox{\plotpoint}}
\multiput(186,75)(16.207,12.966){0}{\usebox{\plotpoint}}
\put(192.76,80.18){\usebox{\plotpoint}}
\multiput(197,83)(17.798,10.679){0}{\usebox{\plotpoint}}
\multiput(202,86)(16.207,12.966){0}{\usebox{\plotpoint}}
\put(209.94,91.76){\usebox{\plotpoint}}
\multiput(212,93)(16.207,12.966){0}{\usebox{\plotpoint}}
\multiput(217,97)(16.207,12.966){0}{\usebox{\plotpoint}}
\put(226.96,103.48){\usebox{\plotpoint}}
\multiput(228,104)(16.207,12.966){0}{\usebox{\plotpoint}}
\multiput(233,108)(16.207,12.966){0}{\usebox{\plotpoint}}
\multiput(238,112)(17.798,10.679){0}{\usebox{\plotpoint}}
\put(243.75,115.60){\usebox{\plotpoint}}
\multiput(248,119)(17.798,10.679){0}{\usebox{\plotpoint}}
\multiput(253,122)(16.207,12.966){0}{\usebox{\plotpoint}}
\put(260.56,127.71){\usebox{\plotpoint}}
\multiput(264,130)(17.798,10.679){0}{\usebox{\plotpoint}}
\multiput(269,133)(16.207,12.966){0}{\usebox{\plotpoint}}
\put(277.42,139.74){\usebox{\plotpoint}}
\multiput(279,141)(17.798,10.679){0}{\usebox{\plotpoint}}
\multiput(284,144)(16.207,12.966){0}{\usebox{\plotpoint}}
\multiput(289,148)(17.798,10.679){0}{\usebox{\plotpoint}}
\put(294.56,151.37){\usebox{\plotpoint}}
\multiput(300,155)(16.207,12.966){0}{\usebox{\plotpoint}}
\multiput(305,159)(17.798,10.679){0}{\usebox{\plotpoint}}
\put(311.55,163.24){\usebox{\plotpoint}}
\multiput(315,166)(16.207,12.966){0}{\usebox{\plotpoint}}
\multiput(320,170)(17.798,10.679){0}{\usebox{\plotpoint}}
\put(328.41,175.28){\usebox{\plotpoint}}
\multiput(331,177)(17.798,10.679){0}{\usebox{\plotpoint}}
\multiput(336,180)(16.207,12.966){0}{\usebox{\plotpoint}}
\put(345.23,187.38){\usebox{\plotpoint}}
\multiput(346,188)(17.798,10.679){0}{\usebox{\plotpoint}}
\multiput(351,191)(16.207,12.966){0}{\usebox{\plotpoint}}
\multiput(356,195)(16.207,12.966){0}{\usebox{\plotpoint}}
\put(362.01,199.50){\usebox{\plotpoint}}
\multiput(367,202)(16.207,12.966){0}{\usebox{\plotpoint}}
\multiput(372,206)(17.798,10.679){0}{\usebox{\plotpoint}}
\put(379.30,210.84){\usebox{\plotpoint}}
\multiput(382,213)(16.207,12.966){0}{\usebox{\plotpoint}}
\multiput(387,217)(17.798,10.679){0}{\usebox{\plotpoint}}
\put(396.21,222.81){\usebox{\plotpoint}}
\multiput(398,224)(16.207,12.966){0}{\usebox{\plotpoint}}
\multiput(403,228)(17.798,10.679){0}{\usebox{\plotpoint}}
\put(412.97,234.98){\usebox{\plotpoint}}
\multiput(413,235)(17.798,10.679){0}{\usebox{\plotpoint}}
\multiput(418,238)(16.207,12.966){0}{\usebox{\plotpoint}}
\multiput(423,242)(16.207,12.966){0}{\usebox{\plotpoint}}
\put(429.86,246.93){\usebox{\plotpoint}}
\multiput(434,249)(16.207,12.966){0}{\usebox{\plotpoint}}
\multiput(439,253)(16.207,12.966){0}{\usebox{\plotpoint}}
\put(446.85,258.71){\usebox{\plotpoint}}
\multiput(449,260)(16.207,12.966){0}{\usebox{\plotpoint}}
\multiput(454,264)(17.798,10.679){0}{\usebox{\plotpoint}}
\put(463.70,270.76){\usebox{\plotpoint}}
\multiput(464,271)(17.270,11.513){0}{\usebox{\plotpoint}}
\multiput(470,275)(17.798,10.679){0}{\usebox{\plotpoint}}
\multiput(475,278)(16.207,12.966){0}{\usebox{\plotpoint}}
\put(480.72,282.58){\usebox{\plotpoint}}
\multiput(485,286)(17.798,10.679){0}{\usebox{\plotpoint}}
\multiput(490,289)(16.207,12.966){0}{\usebox{\plotpoint}}
\put(497.72,294.36){\usebox{\plotpoint}}
\multiput(501,296)(16.207,12.966){0}{\usebox{\plotpoint}}
\multiput(506,300)(16.207,12.966){0}{\usebox{\plotpoint}}
\put(514.67,306.20){\usebox{\plotpoint}}
\multiput(516,307)(16.207,12.966){0}{\usebox{\plotpoint}}
\multiput(521,311)(16.207,12.966){0}{\usebox{\plotpoint}}
\multiput(526,315)(17.798,10.679){0}{\usebox{\plotpoint}}
\put(531.47,318.32){\usebox{\plotpoint}}
\multiput(537,322)(17.798,10.679){0}{\usebox{\plotpoint}}
\multiput(542,325)(16.207,12.966){0}{\usebox{\plotpoint}}
\put(548.47,330.17){\usebox{\plotpoint}}
\multiput(552,333)(17.798,10.679){0}{\usebox{\plotpoint}}
\multiput(557,336)(16.207,12.966){0}{\usebox{\plotpoint}}
\put(565.33,342.22){\usebox{\plotpoint}}
\multiput(568,344)(17.798,10.679){0}{\usebox{\plotpoint}}
\multiput(573,347)(16.207,12.966){0}{\usebox{\plotpoint}}
\put(582.55,353.73){\usebox{\plotpoint}}
\multiput(583,354)(16.207,12.966){0}{\usebox{\plotpoint}}
\multiput(588,358)(16.207,12.966){0}{\usebox{\plotpoint}}
\multiput(593,362)(17.798,10.679){0}{\usebox{\plotpoint}}
\put(599.33,365.89){\usebox{\plotpoint}}
\multiput(604,369)(16.207,12.966){0}{\usebox{\plotpoint}}
\multiput(609,373)(17.798,10.679){0}{\usebox{\plotpoint}}
\put(616.27,377.82){\usebox{\plotpoint}}
\multiput(619,380)(17.798,10.679){0}{\usebox{\plotpoint}}
\multiput(624,383)(16.207,12.966){0}{\usebox{\plotpoint}}
\put(632.92,390.14){\usebox{\plotpoint}}
\multiput(634,391)(18.564,9.282){0}{\usebox{\plotpoint}}
\multiput(640,394)(16.207,12.966){0}{\usebox{\plotpoint}}
\put(649.89,401.91){\usebox{\plotpoint}}
\multiput(650,402)(17.798,10.679){0}{\usebox{\plotpoint}}
\multiput(655,405)(16.207,12.966){0}{\usebox{\plotpoint}}
\multiput(660,409)(17.798,10.679){0}{\usebox{\plotpoint}}
\put(667.12,413.42){\usebox{\plotpoint}}
\multiput(671,416)(16.207,12.966){0}{\usebox{\plotpoint}}
\multiput(676,420)(17.798,10.679){0}{\usebox{\plotpoint}}
\put(684.02,425.41){\usebox{\plotpoint}}
\put(686,427){\usebox{\plotpoint}}
\end{picture} } \]
\caption{A possible choice for the function $f_n$.}\label{fig:fn}
\end{figure}
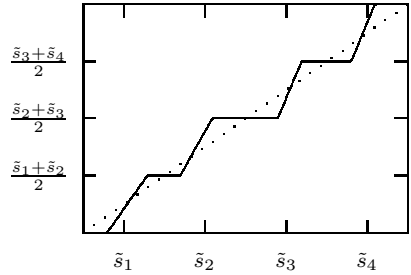

Furthermore $\spec x_{n+1}=s_n+\spec y_n$ so that
$B_{\varepsilon_n}(s_n)$ is in the resolvent set of $x_n$.
The rate of approximation is chosen just so that these gaps remain open
(although possibly become smaller) in every step.
Now we construct an approximation with finite spectrum for each
$x_{n+1}$.
For this we arrange, for fixed $n$, the $s_i,1\leq i\leq n$ into
increasing order, say $\tilde s_1<\ldots<\tilde s_n$, and set
$\delta_n:=\max\{\tilde s_{i+1}-\tilde s_i\mid 1\leq i<n\}$.
We define a continuous monotonically increasing function $f_n$ by
\[ f_n(\lambda) := \begin{cases}
\frac{\tilde s_{i+1}+\tilde s_i}2,&\text{ if }\lambda\in(\tilde
s_i,\tilde s_{i+1})\cap\spec x_{n+1}, \\\text{cont.\ m.~i.}&\text{
else},
 \end{cases} \]
such that $z_{n+1} :=f_n(x_{n+1})\in{\cal A}_{sa}$  (see
figure~\ref{fig:fn}).
 $f_n$ compresses the spectrum between two gaps into one point.
The spectral projections
\begin{align*}
p_i&=P_{[\tilde s_i,\tilde s_{i+1}]}(x_{n+1}),0\leq i\leq n,\text{
where} \\
\tilde s_0&=-\|x_{n+1}\|-1, \\ \tilde s_{n+1}&=\|x_{n+1}\|+1,
\end{align*}
belong to intervals with endpoints in the resolvent set so that they
are in $\cal A$. Therefore
\[ z_{n+1} = \sum_{i=0}^n \frac{\tilde s_i+\tilde s_{i+1}}2 p_i. \]
Thus $\spec z_{n+1}\subset\left\{\frac{\tilde s_i+\tilde
s_{i+1}}2\mid0\leq i\leq n\right\}$, and
with the spectral family $P(\lambda)$ of $x_{n+1}$ we get
\begin{align*} \|z_{n+1}-x_{n+1}\|&\leq  \int |\lambda-f_n(\lambda)|\,
dP(\lambda)\\ &\leq 2\delta_n \|x_{n+1}\| \\
\Rightarrow  \|z_{n+1}-x\|&\leq 2\delta_n\|x_{n+1}\|+\varepsilon.
\end{align*}
Since $\delta_n\rightarrow0$ for $n\rightarrow\infty$ and $\|x_{n+1}\|$
is bounded, we can make the approximation arbitrarily good.
 \end{description}
\end{proof}

\begin{remark}[Kadison property and $RRI_0$]
\label{remark:Kadison property and RRI0}
\begin{enumerate}
\item Kadison property and  property $RRI_0$ are mutually exclusive
since the first forbids existence of projections with
arbitrarily small trace whereas the latter requires this.

\item \CS-algebras  $\cal A$ with $RRI_0$ can contain operators with
band structure: If $\cal A$ is the irrational rotation algebra
(see below) then $\cal A$ has $RRI_0$ by
Theorem~\ref{theorem:properties of the rotation algebra}. But
$\cal A$ contains a subalgebra isomorphic to $C(S^1)$, consisting
of operators with band spectrum only.

\item On the other hand, a \CS-algebra $\cal A$ with the  Kadison
property cannot contain self-adjoint elements with Cantor spectrum:
If $x\in{\cal A}_{sa}$ has Cantor spectrum then every point in $\spec x$
is an accumulation point of $\spec x$ and  $\R\setminus\spec x$, so
that $x$ has no band spectrum in contradiction to
Theorem~\ref{theorem:band structure}.

\item If ${\cal A}_1$ has the Kadison property and ${\cal A}_2$ has
property $RRI_0$ then ${\cal A}:={\cal A}_1\oplus{\cal A}_2$ has
neither of these properties.
\end{enumerate}

\end{remark}

\begin{remark}[real rank and dimension]
\begin{enumerate}
\item If ${\cal A}$ is commutative so that ${\cal A}=C(X)$ for a
topological space $X$ then $\RR({\cal A})=\dim X$ with the usual
definition of dimension.
\item Therefore, \CS-algebras with real rank 0 are (non-commutative)
zero-dimensional spaces.
This includes finite discrete spaces.
However, the additional trace condition in
Theorem~\ref{theorem:Cantor spectrum} excludes finite spaces:
By the Riesz-Kakutani theorem every state on $C(X)$ is given by an
integral with respect to a normalized measure $\mu$,   i.e.\ $\Phi(f)=\int
f\,d\mu$ and  $\mu(X)=1$.
Such states are faithful if and only if every open set has strictly
positive measure.
The trace condition requires that $X$ has connected components with
arbitrary small measure.
\item Every \WS-algebra has real rank 0, since the measurable
functional calculus (as opposed to the continuous) allows to `cut out'
points from the spectrum arbitrarily close.
\item Property  $\RR_0$ is preserved under inductive limits, in
particular  ${\cal A}\otimes{\cal K}$ has real rank 0 if $\RR({\cal
A})=0$.
\end{enumerate}
\end{remark}

\begin{example}[rotation algebra]
The rotation algebra $\mathcal A_\theta$ is the \CS-algebra
generated by two unitaries $U,V$ and the relation
\[ VU=e^{2\pi\imath\theta}UV \] for a given $\theta\in\R$. It also
arises as a reduced twisted group \CS-algebra
$C^*_r(\Z_2,\Theta)$ for the cocycle $\Theta$ given by
$e^{2\pi\imath\theta}$ since $H^2(\Z^2,S^1)\simeq S^1$. It
carries a canonical trace defined by
\[ \tau(1)=1,\tau(U)=\tau(V)=0. \]

The properties of this algebra depend strongly on the nature of
$\theta$:
\begin{theorem}[properties of the rotation
algebra]\label{theorem:properties of the rotation algebra}
\begin{enumerate}
\item If $\theta=p/q$ with $p\in\Z,q\in\N$ co-prime then the
Kadison constant of $\mathcal{A}_\theta$ and of
$\mathcal{A}_\theta\otimes\mathcal{K}$ is $1/q$.
\item If $\theta$ is irrational then $\mathcal{A}_\theta$ and $\mathcal{A}_\theta\otimes\mathcal{K}$ (together
with the canonical trace) have real rank $0$ with infinitesimal
state.
\end{enumerate}
\end{theorem}
\begin{proof}
\begin{enumerate}
\item As is well known, the spectrum of $\mathcal A_\theta$ is $T^2$, all
irreducible representations $\pi_z$ have dimension $q$. The
canonical trace of $a\in\mathcal A_\theta$ is
\[ \tau(a)=\frac1q \int_{T^2}\tr\pi_z(a)\,dz \]
with the canonical trace $\tr$ on $M(q,\C)$. Minimal projections
have rank $1$ in the fiber, and so the Kadison constant is $1/q$.
\item $\mathcal A_\theta$ has real rank zero by
\cite{EllEva:SIRCA}. Since $\mathcal A_\theta$ is simple and
non-elementary also we get $RRI_0$ from \cite[Corollary 8]
{ChoEll:DSAEFSIRCA}.
\end{enumerate}
\end{proof}
\end{example}

\begin{theorem}[Cantor spectrum]
Assume the \CS-algebra ${\cal K}_{\cal A}({\cal E})$ has real rank
$0$ with infinitesimal state. Then every self-adjoint ${\cal
A}$-elliptic operator can be approximated arbitrarily close in
norm resolvent sense by a self-adjoint operator with  Cantor
spectrum.
\end{theorem}
\begin{proof}
Lemma~\ref{lemma:spectral projections} and
Theorem~\ref{theorem:Cantor spectrum}
\end{proof}


\section{Applications} \label{sec:A}

\subsection*{Discrete models}
\begin{example}[generalized Harper operators]
\cite{Sun:DAPMSO} defines magnetic Schr\"o\-din\-ger Operators on
graphs: Let  $X$ be a connected locally finite graph, $\chi$ a
$\C^\times$-valued (i.e.\ non-vanishing complex-valued) map
 (a weight) on the oriented edges $E(X)$, $o,t:E(X)\rightarrow X$ the origin and termination point mappings.
We define a symmetric operator on $l^2(X)$ by
\[ (H_\chi f)(x) = \sum_{\substack{e\in E(X)\\o(e)=x}} \chi(e)f(t(e)) \]
for $f\in l^2(X)$. Two weights $\chi_1,\chi_2$ are called \emph{cohomologous} if there is a
function $s:X\rightarrow S^1$ with
\[ \chi_1(e)=\chi_2(e)\frac{s(o(e))}{s(t(e))} \]
for $e\in E(X)$.

Furthermore, let $\Gamma$ be a group with a properly discontinuous free action on $X$ and such that
the quotient graph is finite (say $n$ points).
A weight  $\chi$ is called  \emph{gauge-invariant} if $\gamma^*\chi$ is cohomologous to $\chi$ for all $\gamma\in\Gamma$.
Then $\chi$ defines a cocycle $\Theta\in Z^2(\Gamma,S^1)$ such that $H_\chi$ commutes with the corresponding
twisted right translations ($R^\Theta_\gamma f(\gamma')=\Theta(\gamma',\gamma)f(\gamma'\gamma)$).
\citeauthor{Sun:DAPMSO} constructs an injective  *-homomorphismus
\[ C^*_r(\Gamma,\Theta)\otimes M(n,\C)\rightarrow \End(l^2(X)), \]
whose image contains $H_\chi$. On the other hand,
\[ {\cal A}\otimes M(n,\C)={\cal K}_{{\cal A}}({\cal A}\otimes\C^n) \]
for the  Hilbert ${\cal A}$-module ${\cal A}\otimes\C^n$ which is the tensor product of the canonical
 module ${\cal A}=C^*_r(\Gamma,\Theta)$ and the Hilbert $\C$-module $\C^n$.
As in Theorem~\ref{theorem:band structure} \citeauthor{Sun:DAPMSO} proves band structure.

All spectral characterizations of this section apply as soon as the corresponding \CS-algebra $C^*_r(\Gamma,\Theta)\otimes M(n,\C)$
fulfills the corresponding assumptions.

We get the ordinary Harper operator for  $E(X)=\Gamma=\Z^2$ and a
suitable graph  $X$ with coordination number $4$ (square lattice),
the hexagonal Harper operator and the quantum pendulum for graphs
with coordination numbers $6$ resp.\ $8$. The corresponding
\CS-algebras are rotation algebras, so that we have band
structure for rational flux, and weak
 genericity of Cantor spectrum for irrational flux.
\end{example}

\subsection*{Continuous models}
\begin{example}[gauge-periodic elliptic operators]
In this case  $\cal A$ will be a twisted group \CS-algebra (left translations), and the Hilbert module will be a
tensor product ${\cal E}={\cal A}\otimes{\cal H}$ with a Hilbert space ${\cal H}$ such that
${\cal K}_{\cal A}({\cal E})\simeq{\cal A}\otimes{\cal K}({\cal H})$.
The operator $D$ will be a differential operator which is invariant under a projective representation of a group,
such as Schr\"odinger, Dirac and Pauli operators with periodic magnetic and electric fields.
\end{example}

The geometric situation we consider is  similar to the case of abelian periodic operators (see
Definition~\ref{definition:periodic operator}) from the introductory section.
Now we allow the group to be non-commutative, and we allow the action to be represented projectivly only on the bundle:

\begin{defiprop}[gauge-periodic operator]
Let $X$ be a smooth ori\-en\-ted Riemannian manifold without
boundary, $\Gamma$ a discrete group acting on $X$ from the left
freely, isometrically, and properly discontinuously. Furthermore,
we assume the action to be cocompact in the sense that the
quotient $M:=X/\Gamma$ is compact. This defines, as in the
abelian case, a left action of $\gamma\in\Gamma$ on smooth
functions $f\in C^\infty(X)$ by
\begin{align}
\gamma^\ma f(x) &:= f(\gamma^{-1}x)
\end{align}
for $x\in X$. As before, this extends to a unitary action on $L^2(X)$.

Next, let $E$ be a smooth Hermitian vector bundle over $X$.
Let  $U$ be a projective representation of  $\Gamma$ in the unitary operators $\mathcal{U}(L^2(E))$ in the following sense:
\begin{gather}
\forall{ \gamma_1,\gamma_2\in\Gamma}:\exists \Theta(\gamma_1,\gamma_2)\in C(X,S^1):
U_{\gamma_1}U_{\gamma_2}=\Theta(\gamma_1,\gamma_2)U_{\gamma_1\gamma_2}. \label{e:PraeKozyklus}
\end{gather}
Assume that $U$ is a (projective) lift of the $\Gamma$-action on $C^\infty(X)$, i.e.
\begin{equation}
\forall{ \varphi\in C^\infty_c(X)}:\forall{s\in L^2(E)}:\forall{\gamma\in\Gamma}:
U_\gamma(\varphi s)=(\gamma^\ma \varphi)U_\gamma(s).
\end{equation}
Assume that $U$ is smooth, i.e.
 $\forall{ \gamma\in\Gamma}: U_\gamma(C^\infty(E)\cap L^2(E)) \subset C^\infty(E)$.
Then  $U_\gamma$ is $\gamma$-local, i.e.
\begin{equation} \forall{ s\in C^\infty(E)}: \supp (U_\gamma s)\subset \gamma\supp s,\end{equation}
and it leaves the domain  $\mathcal{D}(D)=C^\infty_c(E)$ of any differential operator $D$ on $E$ invariant.
We call $D$  \textbf{gauge-periodic} if, on  $\mathcal{D}(D)$, one has:
\begin{equation} \forall{\gamma\in\Gamma}: [U_\gamma,D]=0 \end{equation}
\end{defiprop}
\begin{proof} Let $x\in X\setminus\supp s$. Since $\supp s$ is closed there is a neigborhood $O\subset X$ of $x$
and $\varphi\in C^\infty_c(X)$ with $\varphi|O=1$, $\varphi|_{\supp s}=0$.
Then  $(1-\varphi) s=s$ and therefore
\begin{equation*}  \begin{split} U_\gamma s &= U_\gamma\left((1-\varphi)s\right) \\
  &= (1-\gamma^\ma \varphi)U_\gamma s \\
  &= 0 \text{ on } \gamma O.
\end{split} \end{equation*}
Since $U$ is smooth also, it leaves $C^\infty_c(E)$ invariant.
\end{proof}

\begin{proposition}[cocycle property] $\Theta$ fulfills the cocycle property:
\begin{equation} \forall{ \gamma_1,\gamma_2,\gamma_3\in\Gamma}:
\Theta(\gamma_1,\gamma_2)\Theta(\gamma_1\gamma_2,\gamma_3)=
 \Theta(\gamma_1,\gamma_2\gamma_3)\gamma_1^\ma[\Theta(\gamma_2,\gamma_3)]\label{e:Kozyklus}
\end{equation}
\end{proposition}
\begin{proof} This follows from associativity $U_{\gamma_1}(U_{\gamma_2}U_{\gamma_3})=
(U_{\gamma_1}U_{\gamma_2})U_{\gamma_3}$ and the projectivity condition~\eqref{e:PraeKozyklus}:
\begin{equation*} \begin{split}
U_{\gamma_1}(U_{\gamma_2}U_{\gamma_3}) &=  U_{\gamma_1}\Theta(\gamma_2,\gamma_3)U_{\gamma_2\gamma_3} \\
&= \gamma_1^\ma[\Theta(\gamma_2,\gamma_3)] U_{\gamma_1}U_{\gamma_2\gamma_3} \\
                                       &= \Theta(\gamma_1,\gamma_2\gamma_3)\gamma_1^\ma[\Theta(\gamma_2,\gamma_3)]
                                       U_{\gamma_1\gamma_2\gamma_3} \\
(U_{\gamma_1}U_{\gamma_2})U_{\gamma_3} &= \Theta(\gamma_1,\gamma_2)U_{\gamma_1\gamma_2}U_{\gamma_3} \\
                                       &= \Theta(\gamma_1,\gamma_2)\Theta(\gamma_1\gamma_2,\gamma_3)
                                       U_{\gamma_1\gamma_2\gamma_3}
\end{split} \end{equation*}
\end{proof}

\begin{remark}[exact cocycle and representation]\label{b:Theta=1}
$\Theta$ therefore defines a class in the group cohomology $H^2(\Gamma,C(X,S^1))$
\cite[see, e.g.,][]{Bro:CG}. Exact 2-cocycles have the form
\begin{equation} \Theta(\gamma,\gamma')=\sigma(\gamma)\gamma^\ma[\sigma(\gamma')]\sigma(\gamma\gamma')^{-1} \end{equation}
with a 1-cocycle $\sigma$, so they define a proper representation of $\Gamma$ by
\begin{equation} \tilde U_\gamma := \sigma(\gamma)^{-1}U_\gamma, \end{equation}
which also commutes with $D$ if the cocycle is constant in $x\in X$.
Without loss of generality we assume that  $\Theta$ is normalized, i.e.\ $\Theta(e,e)=1$.
\end{remark}

\begin{proposition}[bundle morphisms]\label{proposition:bundle morphisms}
$U$ defines a family  $u$ of vector bundle morphisms on $E$,  $u_\gamma:E_x\rightarrow E_{\gamma x}$. $u$ is a
 projective lift of the $\Gamma$-action from $X$ to $E$,  i.e.\
\begin{equation} \forall{ \gamma_1\gamma_2\in\Gamma}: u_{\gamma_1}u_{\gamma_2}=
\Theta(\gamma_1,\gamma_2)u_{\gamma_1\gamma_2} \end{equation}
with the same cocycle $\Theta$ as for $U$. $u$ induces $U$ via
\begin{equation} (U_\gamma s)(x):= u_\gamma s(\gamma^{-1}x). \end{equation}
If $t$ is a (proper) lift of the  $\Gamma$-action from $X$ to $E$ and $T$ the induced action
\begin{equation} (T_\gamma s)(x):= t_\gamma s(\gamma^{-1}x) \end{equation}
on $C^\infty(X)$ then  $u$ and $U$ can be expressed as  $u=mt$ and $U=MT$, where $m$ is a family of (strict)
vector bundle isomorphisms.
\end{proposition}
\begin{proof}
Let $v\in E_x$. We choose  $s\in C^\infty(E)$ with $s(x)=v$ and
set -- a priori depending on $s$ -- $u^s_\gamma(v):=
\left(U_\gamma(s)\right)(\gamma x)\in E_{\gamma x}$. If
$\varphi\in C^\infty(x)$, $\varphi(x)=1$, we get
\begin{equation} \begin{split} u^{\varphi s}_\gamma(v) & =
(\gamma^\ma \varphi)(\gamma x)\left(U_\gamma(s)\right)(\gamma x) \\
 & = u^s_\gamma(v), \end{split} \end{equation}
 i.e.\ $u^s_\gamma(v)$ depends on the value of  $s$ at the point  $x$ only;
hence we omit $s$ in the notation.
The morphism property follows from the corresponding property of  $U_\gamma$, and from $(u_\gamma)^{-1}=u_{\gamma^{-1}}$.

$u$ induces $U$ by construction.

If there is a proper lift $t$ then  $m:=ut^{-1}$ defines the strict morphism we look for:
\[
\begin{CD}
 E @>t_\gamma^{-1}>> E @>u_\gamma>> E \\
 @VVV @VVV @VVV \\
 X @>\gamma^{-1}>> X @>\gamma>> X
\end{CD}
\]
\end{proof}

\begin{remark}[lift of the action]
If  $\Theta$ is exact and $\tilde u$ the family of vector bundle isomorphisms belonging to $\tilde U$ by
remark~\ref{b:Theta=1} then  $\tilde u$ is a proper lift of the $\Gamma$-action from $X$ to $E$.
\end{remark}

\begin{proposition}[properties of the cocycle]
\begin{enumerate}
\item $\forall{\gamma\in\Gamma}:\Theta(\gamma,e)=\Theta(e,\gamma)=1$
\item $\forall{\gamma\in\Gamma}:\Theta(\gamma,\gamma^{-1})=\Theta(\gamma^{-1},\gamma)$
\end{enumerate}
\end{proposition}
\begin{proof} Easy consequences of the cocycle property.
\end{proof}


For the case of a bicharacter $\Theta$ \cite{BruSun:SGPEO,BruSun:TGCASGPO} describe how to construct a parametrix
 for elliptic gauge-periodic differential operator by lifting and translating a parametrix for a fundamental domain.
  The same construction works for the slightly more general case of a 2-cocycle.

From this one concludes as in the cited work:
\begin{theorem}[self-adjointness]\label{theorem:GPMO}
Every symmetric elliptic gauge-periodic differential operator is essentially self-adjoint on $C_c^\infty(E)$.
\end{theorem}

Similarly, a trivial extension of \cite{BruSun:SGPEO,BruSun:TGCASGPO} shows how to construct the heat kernel:
\begin{theorem}[heat kernel]\label{theorem:heat kernel}
Let $D$ be a symmetric elliptic gauge-periodic differential operator, bounded below, of order $p>d=\dim X$.
Then  $e^{-t\overline D}$ has, for $t>0$, a smooth integral kernel $K_t(x,y)\in E_x\otimes E_y^*$ such that
\begin{equation}
|K_t(x,y)|\leq C_1 t^{-d/p}\exp\left( -C_2\dist(x,y)^{p/(p-1)}t^{-1/(p-1)} \right)
\end{equation}
with $C_1,C_2>0$, uniformly on $(0,T]\times X\times X$.
\end{theorem}

Again following
 \cite{BruSun:SGPEO} we construct a suitable decomposition of $L^2(E)$.
For that we choose a fundamental domain $\cal D$ for the $\Gamma$-action, set ${\cal H}=L^2\left(E|_{\cal D}\right)$
and define a unitary map by
\begin{align*}
\Phi:L^2(E)&\rightarrow l^2(\Gamma,{\cal H})\simeq l^2(\Gamma)\otimes{\cal H},\\
 \Phi(s)(\gamma)&= (U_\gamma (s))|_{\cal D}.
\end{align*}
Then we have for $f\in l^2(\Gamma)\otimes{\cal H}$
\begin{align*}
(\Phi U_\gamma\Phi^* f)(\gamma') &= (U_{\gamma'}U_\gamma\Phi^* f)|_{\cal D}\\
 &= \Theta(\gamma',\gamma)(U_{\gamma'\gamma}\Phi^* f)|_{\cal D}\\
 &= \Theta(\gamma',\gamma)(\Phi\Phi^* f)(\gamma'\gamma) \\
 &= \Theta(\gamma',\gamma)f(\gamma'\gamma) \\
 &=: \Theta(\gamma',\gamma) R_\gamma f(\gamma') = R^\Theta_\gamma f(\gamma')
\end{align*}
with the right translation $R_\gamma$ and twisted right translation $R^\Theta_\gamma f(\gamma')$.

So it's natural to try and define a  $C^*_r(\Gamma,\Theta)$-action on $L^2(E)$ by
\[ R^\Theta_\gamma(s) = U_\gamma(s) \]
for $s\in L^2(E)$.
Here, the cocycle  $\Theta$ can in general depend on  $x\in X$ so that we have to find the gauge-translations
in  $C(X,S^1)\times_{\alpha,\theta}\Gamma$.
This  \CS-algebra has interesting structural properties but is not suitable for the applications on spectral theory
 developed in the previous section.

If $\Theta$ is periodic in $x\in X$ then we get a field of twisted reduced group \CS-algebras  $C^*_r(\Gamma,\Theta_x),x\in M$ over $M$.
In general this field is still to `large'.

Therefore we require the cocycle to be constant in $x\in X$, so that we have to deal with the reduced twisted
group \CS-algebra $C^*_r(\Gamma,\Theta)$ only.
This is still general enough for the applications we are interested in: magnetic Schr\"odinger operators (and there
 Pauli and Dirac analogs).

Now note that $l^2(\Gamma)$ is the GNS representation space of ${\cal A}:=C^*_r(\Gamma,\Theta)$ with respect to the
 canonical trace given by
\begin{align*}
\tau\left(R^\Theta_\gamma\right) &=
\begin{cases}
1,&\gamma=e, \\
0,else,
\end{cases}
\end{align*}
and that $l^1(\Gamma)\subset C^*_r(\Gamma,\Theta)\subset l^2(\Gamma)$.
Therefore it's natural to view the left Hilbert-$\cal A$ module\footnote{The
action is naturally a left action since it is given by endomorphisms on a vector space.}
as ${\cal E}:={\cal H}\otimes{\cal A}$ so that $L^2(E)$ is the Hilbert-GNS representation space of ${\cal E}$.
To define the scalar product we use the observations made in the commutative case (see Definition and
Proposition~\ref{defiprop:mit dem bloeden Label}).

\begin{lemma}[left pre-Hilbert ${\cal A}$-module]\label{lemma:left pre-Hilbert A-module}
\begin{align}
\<s_1|s_2>_{\cal E} &= \sum_{\gamma\in\Gamma}\<U_{\gamma}s_2|s_1>_{L^2(E)} R^\Theta_{\gamma} \label{equ:A-Modulstruktur}
\end{align}
for $s_1,s_2\in C_c(E)$ defines the structure of a left pre-Hilbert ${\cal A}$-module on $C_c(E)$; under the isomorphism
 $\Phi$ it coincides with the left tensor Hilbert ${\cal A}$-module structure of ${\cal H}\otimes{\cal A}$.
\end{lemma}
\begin{proof}
For $f_1,f_2\in{\cal H},a_1,a_2\in{\cal A}$ we have by definition
\[ \<a_1\otimes f_1|a_2\otimes f_2>_{{\cal A}\otimes{\cal H}} = \<f_2|f_1>_{\cal H}a_1a_2^*, \]
since a left  Hilbert-$\C$-module is a Hilbert space with conjugated scalar product (complex linear in the first argument,
 anti-linear in the second).
For $s_1,s_2\in C_c(E)$ we get after identifying $\delta_{\gamma^{-1}}$ with
$\bar\Theta(\gamma,\gamma^{-1})R^\Theta_\gamma$
{\allowdisplaybreaks
\begin{multline*}
\<\Phi(s_1)|\Phi(s_2)>_{{\cal A}\otimes{\cal H}} = \sum_{\gamma,\gamma'\in\Gamma}\<\delta_\gamma\otimes\Phi(s_1)
(\gamma)|\delta_{\gamma'}\otimes\Phi(s_2)(\gamma')>_{{\cal A}\otimes{\cal H}} \\
\begin{split}
&= \sum_{\gamma,\gamma'\in\Gamma}\<\Phi(s_2)(\gamma')|\Phi(s_1)(\gamma)>_{{\cal H}} \bar\Theta(\gamma^{-1},\gamma)
\Theta(\gamma'{}^{-1},\gamma')R^\Theta_{\gamma^{-1}} \left(R^\Theta_{\gamma'{}^{-1}}\right)^{*}\\
&= \sum_{\gamma,\gamma'\in\Gamma}\<\Phi(s_2)(\gamma')|\Phi(s_1)(\gamma)>_{{\cal H}} \bar\Theta(\gamma^{-1},\gamma)
R^\Theta_{\gamma^{-1}} R^\Theta_{\gamma'}\\
&= \sum_{\gamma,\gamma'\in\Gamma}\<(U_{\gamma'}s_2){|_{\cal D}}|(U_\gamma s_1){|_{\cal D}}>_{{\cal H}}
 \bar\Theta(\gamma^{-1},\gamma) \Theta(\gamma^{-1},\gamma')R^\Theta_{\gamma^{-1}\gamma'}\\
&= \sum_{\gamma,\gamma'\in\Gamma}\<(U_{\gamma\gamma'}s_2){|_{\cal D}}|(U_\gamma s_1){|_{\cal D}}>_{{\cal H}}
\bar\Theta(\gamma^{-1},\gamma) \Theta(\gamma^{-1},\gamma\gamma')R^\Theta_{\gamma'}\\
&= \sum_{\gamma,\gamma'\in\Gamma}\<(U_{\gamma}U_{\gamma'}s_2){|_{\cal D}}|(U_\gamma s_1){|_{\cal D}}>_{{\cal H}}
 \Theta(\gamma,\gamma')\bar\Theta(\gamma^{-1},\gamma) \Theta(\gamma^{-1},\gamma\gamma')R^\Theta_{\gamma'}\\
&= \sum_{\gamma,\gamma'\in\Gamma}\<(U_{\gamma}U_{\gamma'}s_2){|_{\cal D}}|(U_\gamma s_1){|_{\cal D}}>_{{\cal H}}
 R^\Theta_{\gamma'}\\
&= \sum_{\gamma'\in\Gamma}\<U_{\gamma'}s_2|s_1>_{L^2(E)} R^\Theta_{\gamma'}.
\end{split}
\end{multline*}
}
This shows that the structures coincide.
\end{proof}

\begin{lemma}[left Hilbert ${\cal A}$-module]
The completion of the left pre-Hilbert ${\cal A}$-module $C_c(E)$ is isomorphic to ${\cal E}={\cal H}\otimes{\cal A}$.
The GNS representation of  ${\cal E}$ with respect to the canonical trace  $\tau$ on $\cal A$ is isomorphic to $L^2(E)$.
\end{lemma}
\begin{proof}
By equation~\eqref{equ:A-Modulstruktur} we have for $s\in C_c(E)$
\begin{align*}
\|s\|^2_{\cal E} &= \|\<s|s>_{\cal E}\|_{\cal A} \\
 &\geq \<s|s>_{L^2(E)}.
\end{align*}
Therefore, the completion of  $C_c(E)$ with respect to $\|\cdot\|_{\cal E}$ is contained in the one with respect
to $\|\cdot\|_{L^2(E)}$, i.e.\ in $L^2(E)$. But  $C_c(E)$ is dense in $\cal E$.

We get the scalar product of the GNS representation with respect to  $\tau$ for  $s_1,s_2\in C_c(E)$ from
\begin{align*}
\<s_1|s_2>_\tau &= \tau\left(\<s_2|s_1>_{\cal E}\right) \\
 &= \tau\left(\sum_{\gamma\in\Gamma}\<U_{\gamma}s_1|s_2>_{L^2(E)} R^\Theta_{\gamma} \right) \\
 &= \sum_{\gamma\in\Gamma}\<U_{\gamma}s_1|s_2>_{L^2(E)} \tau\left(\tilde\rho(\delta_\gamma)\right) \\
 &= \<s_1|s_2>_{L^2(E)}.
\end{align*}
Since $C_c(E)\subset {\cal E}\subset L^2(E)$ is dense the GNS representation space is exactly $L^2(E)$.
\end{proof}

\begin{lemma}[$\cal A$-compact operators]
The $\cal A$-compact operators on  $\cal E$ are given by
\begin{align} {\cal K}_{\cal A}({\cal E})&\simeq {\cal A}^{op}\otimes{\cal K}. \end{align}
Here  $\cal K$ denotes the compact  operators on ${\cal H}=L^2(E|_{\cal D})$, and ${\cal A}^{op}$ is
the  \CS-Algebra $C^*_r(\Gamma,\theta)^L$ generated by the left translations twisted with $\theta$.
\end{lemma}
\begin{proof}
For tensor products of left Hilbert modules we have in general
\begin{align*} {\cal K}_{\cal A}({\cal A}\otimes{\cal H}) &\simeq {\cal K}_{\cal A}({\cal A})\otimes{\cal K}_{\C}({\cal H})\\
&\simeq {\cal A}^{op}\otimes{\cal K}.\end{align*}
The statement about ${\cal A}^{op}$ is well known in the untwisted case since the opposite of left multiplication is right
 multiplication. It is easy to check that this holds in the twisted case also.
\end{proof}

Following our rationale from Section~\ref{sec:NBT} we define a trace $\tr_\tau$ and identify
bounded module operators in ${\cal L}_{\cal A}({\cal E})$ with their images in ${\cal L}(L^2(E))$ under
the faithful representation with respect to  $\tau$.

As in  \cite{BruSun:SPEO,BruSun:TGCASGPO} one shows, using Theorem~\ref{theorem:heat kernel}:
\begin{theorem}[gauge-periodic operators]\label{theorem:gauge-periodic operators}
Let $D$ be a symmetric gauge-periodic differential operator.
The the resolvent of  $\bar D$ is $\cal A$-compact, and $e^{-t\bar D^2}$ is $\tr_\tau$-trace class.
\end{theorem}

\begin{theorem}[gauge-periodic module operators]\label{theorem:gauge-periodic module operators}
Let $D$ be a symmetric gauge-peri\-odic differential operator. Then  $D$ defines an $\cal A$-elliptic operator $T$
such that the resolvents of  $\bar D$ and $\bar T$ coincide (under the GNS representation).
\end{theorem}
\begin{proof}
Set  ${\cal D}(T):={\cal D}(D)=C^\infty_c(E)$.
Then  ${\cal D}(T)\subset{\cal E}$ dense, we set $T:=D$ as operators on vector spaces.

$T$ is adjointable since $D$ is symmetric and gauge-periodic:
For $s_1,s_2\in C^\infty_c(E)$ we have
\begin{align*}
\<s_1|Ds_2>_{\cal E} &= \sum_{\gamma\in\Gamma}\<U_{\gamma}s_1|Ds_2>_{L^2(E)} R^\theta_\gamma \\
 &= \sum_{\gamma\in\Gamma}\<DU_{\gamma}s_1|s_2>_{L^2(E)} R^\theta_\gamma \\
 &= \sum_{\gamma\in\Gamma}\<U_{\gamma}Ds_1|s_2>_{L^2(E)} R^\theta_\gamma \\
 &= \<Ds_1|s_2>_{\cal E}.
\end{align*}
Finally,  $\ran (1+D^*D)$ is dense in $L^2(E)$ because $D$ is essentially self-adjoint; therefore, $T$ is regular.
\end{proof}

This allows to apply all of the spectral characterizations from the previous section.

\begin{example}[periodic elliptic operator]\label{example:PEO}
A gauge-periodic operator is called \emph{periodic} if the corresponding cocycle fulfills $\Theta\equiv1$.
If the group $\Gamma$ is abelian then we are back in the commutative case (see Definition~\ref{definition:periodic operator})
where ordinary Bloch theory applies.
If $\Gamma$ is not abelian then it doesn't apply, although the cocycle is trivial.
But it is still covered by non-commutative Bloch theory, of course.
\end{example}

\begin{example}[magnetic Schr\"odinger operator]
In example~\ref{example:sowepmf} and remark~\ref{remark:nemf} we saw that the magnetic Schr\"dinger operator with a
magnetic field
$b\in\Omega^2(X),db=0,[\frac1{2\pi}b]\in H^2(X,\Z)$ is given by a (symmetric elliptic)
Bochner-Laplace operator on a Hermitian line bundle $L$ over $X$ with curvature $b$.
It is gauge-periodic with possibly non-constant cocycle if $H^1(X,S^1)=0$ (see remark~\ref{remark:nemf} and the work cited
there).
If $b$ is exact then the cocycle can be chosen to be constant.
If the magnetic flux is integral ($b_M\in H^2(M,\Z)$, see example~\ref{example:sowepmf})
then the operator is periodic.
If there is a periodic magnetic potential $a$ for $b=da$ (i.e.\ if the magnetic flux is $0$) then the operator
is strictly periodic in the usual sense of ordinary Bloch theory, i.e.\ it is a periodic operator on $L^2(X)$ (no
magnetic translations, no bundles).
\end{example}

\begin{example}[magnetic Schr\"odinger operator on $\R^2$]
In the Euclidean case, if  $\Gamma=\Z^2$  we end up with a
rotation algebra $\mathcal A_\theta$ where $\theta$ is given by
the magnetic flux. So, from Theorem~\ref{theorem:properties of
the rotation algebra} we get band structure in the case of
rational flux and weak genericity of Cantor spectra in the case
of irrational flux. Since it is a criterion inside the algebra of
symmetries it applies to the corresponding Pauli and Dirac
operators as well.
\end{example}

\begin{example}[magnetic Schr\"odinger operator on $\mathbb H^2$]
To investigate the importance of the geometry it is interesting to study the hyperbolic analog, since the
corresponding cocompact groups (Fuchsian groups) are non-amenable and therefore `opposite' to the amenable
groups in the Euclidean case. The analog of a constant magnetic field is a constant multiple of the volume form.
\cite{CarHanMatMcC:QHEHP,CarHanMat:QHEHPPD} compute $K$-groups and Kadison constants for twisted Fuchsian groups:
Again, one has Kadison property if and only if the magnetic flux is rational.

\cite{MarMat:TITGOINBT,MarMat:THITGOIIFQN} study similar questions for good orbifolds.
\end{example}

\begin{example}[gauge-periodic point perturbations]
In Euclidean space, point perturbations provide explicitly solvable models for periodic Schr\"odinger operators.
\cite{BruGei:GPPPLP,BruGei:SPPPKRF} show how to define these types of operators more generally in our given geometric
context (manifold with cocompact group action). If the point perturbation is gauge-periodic, then the perturbed operator
is gauge-periodic in our sense, so that non-commutative Bloch Theory applies. In particular, periodic point
perturbations of the magnetic Schr\"odinger operator with rational flux have band structure.
\end{example}

\subsection*{Elliptic operators on Hilbert module bundles}
\begin{example}
\cite{FomMis:IEOCA} extended the usual notion of an index of an operator by replacing Hilbert spaces by Hilbert modules:
Let $\cal A$ be a  \CS-algebra, $M$ a compact Riemannian manifold and $E$ a bundle over $M$ of  Hilbert $\cal A$-modules
 (a \emph{{Hilbert module bundle}}).
On can define Sobolev norms as usual, now coming from an  $\cal A$-scalar product.
Thus one gets a scale of Sobolev-Hilbert $\cal A$-modules for which the Sobolev lemma holds.
Instead of the usual pseudo-differential operators whose coefficients are vector space endomorphisms one has
$\cal A$-pseudo-differential operators with coefficients in the bundle
${\cal L}_{\cal A}(E):=\bigcup_{x\in M}{\cal L}_{\cal A}(E_x)$.
They act in the usual way on the  Sobolev-Hilbert modules.
Symbols of $\cal A$-pseudo-differential operators are represented by section of ${\cal L}_{\cal A}(E)$.
As in the scalar case, an elliptic operator has an  $\cal A$-compact resolvent, hence it is $\cal A$-elliptic in
the sense of Definition~\ref{definition:A-elliptic operator}.
Furthermore, elliptic operators are $\cal A$-Fredholm and therefore have an index in $K_0({\cal A})$.

A special case are the periodic elliptic operators:
Let $X$ be a Riemannian manifold with properly discontinuous, isometric, cocompact action of a group $\Gamma$,
 and $D$ a $\Gamma$-periodic operator as in example~\ref{example:PEO}, $M:=\Gamma\backslash X$.
Let  $\rho$ be the right regular representation\footnote{Usually one studies the right regular representation on
the vector space $\C\Gamma\subset C^*_r(\Gamma)$ or on the Hilbert space $l^2(\Gamma)\supset C^*_r(\Gamma)$.
But $C^*_r(\Gamma)$ is a $\Gamma$-invariant subspace of $l^2(\Gamma)$.} of  $\Gamma$ on ${\cal A}:=C^*_r(\Gamma)$.
Then $X\times_\rho {\cal A}$ is an ${\cal A}$-bundle over $M$ on which  $D$ acts.
Besides,  ${\cal A}$ carries the structure of a standard Hilbert-${\cal A}$ module.
If $D$ is elliptic then  $D$ determines an elliptic operator on $X\times_\rho {\cal A}$.
\end{example}


\appendix

\section{Continuous fields of Hilbert spaces}\label{sec:CFHS}
We follow the classic reference \cite{DixDou:CCEHCA}.

\begin{definition}[continuous fields of Banach and Hilbert spaces]
Let $B$ be a topological space, $\left(E(z)\right)_{z\in B}$ a family
of Banach spaces.
The linear space $\Pi:=\prod_{z\in B}E(z)$ is called \emph{space of all
vector fields}.
A \emph{continuity structure} on $\Pi$ is defined by a subspace
$\Lambda\subset\Pi$ such that:
\begin{enumerate}
\item $\Lambda$ is a $C_\infty(B)$-submodule of $\Pi$.
\item $\forall{z\in B}:\forall{\xi\in E(z)}:\exists
x\in\Lambda:x(z)=\xi$
\item $\forall{x\in\Lambda}: \left(z\mapsto\|x(z)\|\right)\in
C_\infty(B)$
\item \label{item:complete} \begin{tabular}[t]{@{}l@{}r@{}}
$\forall{x\in\Pi}: \big\langle$ & $\langle
\forall{\varepsilon>0}:\forall{z\in B}:\exists x'\in\Lambda,
\text{neighborhood }U\ni z: $ \\ & $\forall{z'\in U}: \|x(z')-
x'(z')\|<\varepsilon  \rangle  \Rightarrow x\in\Lambda \big\rangle$
\end{tabular}
\end{enumerate}
${\mathcal E}:=((E(z))_{z\in B},\Lambda)$ is called \emph{continuous
field of Banach spaces}.
If the fibers  $E(z)$ are Hilbert spaces we have a  \emph{continuous
field of Hilbert spaces}.
The scalar product is automatically continuous.
\end{definition}
Condition~\ref{item:complete} is a completeness condition:
If a vector field $x\in\Pi$ can be locally approximated arbitrarily
well by continuous vector fields then it is continuous.

\begin{proposition}[defining submodule]\label{proposition:DS}
Let  $B,\Pi$ be as above and  $\Lambda\subset\Pi$ a subspace with
\begin{enumerate}
\item $\forall{z\in B}:\{x(z)\mid x\in\Lambda\} =: \Lambda(z)$ dense in
$E(z)$ and
\item $\forall{x\in\Lambda}: (z\mapsto\|x\|)\in C_\infty(B)$.
\end{enumerate}
Then there is a unique continuity structure $\tilde\Lambda$ on $\Pi$
with $\tilde\Lambda\supset\Lambda$.
$\tilde\Lambda$ is given by
\begin{multline*} \tilde\Lambda=\{ x\in\Pi\mid \forall{z\in B}:\varepsilon>0\exists
\text{neighborhood }U\ni z,x'\in\bar\Lambda: \\ \forall{z'\in U}:\|x(z')-
x'(z')\|<\varepsilon\}.
\end{multline*}
\end{proposition}

\begin{lemma}[continuous fields and Banach space bundles]
A continuous field of Banach spaces ${\mathcal E}$ defines a Banach
space bundle\footnote{
A bundle has a continuous open surjection onto the base, but is not
necessarily locally trivial.
However, for a locally compact base and finite dimensional fibers this
follows from the existence of the projection.}
 $E$ over $B$ so that the continuous sections $C(E)$ are the continuous
vector fields of ${\mathcal E}$.
\end{lemma}

\begin{proof}[Sketch of the proof]
As a set $E:=\prod_{z\in B}E(z)$. We choose the topology so that the
natural projection $\pi:E\rightarrow B$ is continuous and open:
The topology is generated by the tubular neighborhoods
\[ T(U,x,\varepsilon):=\{ \xi\in E\mid \pi(\xi)\in U \wedge
\|\xi-x(\pi(\xi))\|<\varepsilon \} \]
for open sets $U\subset B$, continuous fields $x\in{\mathcal E}$ and
$\varepsilon>0$ .
It is easy to check that the tubular neighborhoods generate a topology
on $E$ with the desired properties.
On the fibers $E(z)$ it induces the strong topology since the
intersections $E(z)\cap T(U,x,\varepsilon)$ of the fibers with the
tubular neighborhoods are just norm balls in the fiber.
\end{proof}

\begin{lemma}[continuous field as Hilbert \CS-module]\label{lemma:CFHCM}
A continuous field of Hilbert spaces  ${\mathcal E}=((E(z))_{z\in
B},\Lambda)$ over $B$ defines a  Hilbert $C_\infty(B)$-module structure
on $\Lambda$.
Vice versa: A Hilbert $C_\infty(B)$-module defines a continuous field
of Hilbert spaces, and this correspondence is one-to-one.
\end{lemma}

\section{Hilbert \CS-modules}\label{sec:HCM}
Usually Hilbert \CS-modules are defined to be right modules. We define
these and the left modules and list basic properties and objects
connected to them.

\begin{definition}[(right) Hilbert module]\label{definition:HRM}
Let  $\cal A$ be a  \CS-algebra. A right $\cal A$-module ${\cal E}$
is called (right) pre-Hilbert $\cal A$-module if it is endowed with a map
 $\<\cdot|\cdot>:{\cal E}\times {\cal E}\rightarrow{\cal
A}$ with the following properties:
\begin{enumerate}
\item $\<e|f+g>=\<e|f>+\<e|g>$ for $e,f,g\in {\cal E}$.
\item $\<e|f\lambda>=\<e|f>\lambda$ for $e,f\in {\cal E},\lambda\in\C$.
\item $\<e|fa>=\<e|f>a$ for $e,f\in {\cal E},a\in{\cal A}$.
\item  $\<f|e>=\<e|f>^*$ for $e,f\in {\cal E}$. \label{enum:fe=ef*}
\item  $\<e|e>\geq0$ in $\cal A$ for  $e\in {\cal E}$, and
$\<e|e>=0\Leftrightarrow e=0$. \label{enum:e=0}
\end{enumerate}
Then the map ${\cal E}\ni e\mapsto \sqrt{\|\<e|e>\|_{\cal A}}$
defines a norm on  ${\cal E}$. The \emph{closure} of $\cal
E$ is defined as the completion
of $\cal E$ as Banach space with this norm.

${\cal E}$ is called \emph{(right) Hilbert $\cal A$-module} if ${\cal
E}$ is complete in this norm.

An operator  $T\in{\cal L}({\cal E})$ is called \emph{adjointable} if
there is  $T^*\in{\cal L}({\cal E})$ such that for all $e,f\in {\cal
E}$: $\<e|Tf>=\<T^*e|f>$. The set of adjointable operators
is denoted by \emph{${\cal L}_{\cal A}({\cal
E})$}.

For $e,f\in {\cal E}$ we define an operator $\pi_{e,f}$ by
\[ \pi_{e,f}:{\cal E}\ni x\mapsto e\<f|x>\in {\cal E}. \]
We set \emph{${\cal F}_{\cal A}({\cal
E})$}$:=\operatorname{span}\{\pi_{e,f}\mid e,f\in {\cal
E}\}$ and call this the set of
 \emph{$\cal A$-finite operators}. The set
 \emph{${\cal K}_{\cal A}({\cal E})$} of \emph{$\cal A$-compact
operators} is
the closure of   ${\cal F}_{\cal A}({\cal E})$ in ${\cal L}_{\cal
A}({\cal E})$.
\end{definition}
The brackets indicate that by Hilbert module we mean a right Hilbert module.

\begin{definition}[left Hilbert module]\label{definition:HLM}
Let  $\cal A$ be a  \CS-algebra. A left $\cal A$-module ${\cal E}$
is called left pre-Hilbert $\cal A$-module if it is endowed with a map
 $\<\cdot|\cdot>:{\cal E}\times {\cal E}\rightarrow{\cal
A}$ with the following properties:
\begin{enumerate}
\item $\<e+f|g>=\<e|g>+\<f|g>$ for $e,f,g\in {\cal E}$.
\item $\<\lambda e|f>=\lambda\<e|f>$ for $e,f\in {\cal E},\lambda\in\C$.
\item $\<ae|f>=a\<e|f>$ for $e,f\in {\cal E},a\in{\cal A}$.
\item  $\<f|e>=\<e|f>^*$ for $e,f\in {\cal E}$.
\item  $\<e|e>\geq0$ in $\cal A$ for  $e\in {\cal E}$, and
$\<e|e>=0\Leftrightarrow e=0$.
\end{enumerate}
Then the map ${\cal E}\ni e\mapsto \sqrt{\|\<e|e>\|_{\cal A}}$
defines a norm on  ${\cal E}$. The \emph{closure} of $\cal
E$ is defined as the completion
of $\cal E$ as Banach space with this norm.

${\cal E}$ is called \emph{left Hilbert $\cal A$-module} if ${\cal
E}$ is complete in this norm.

An operator  $T\in{\cal L}({\cal E})$ is called \emph{adjointable} if
there is  $T^*\in{\cal L}({\cal E})$ such that for all $e,f\in {\cal
E}$: $\<e|Tf>=\<T^*e|f>$. The set of adjointable operators
is denoted by \emph{${\cal L}_{\cal A}({\cal
E})$}.

For $e,f\in {\cal E}$ we define an operator $\pi^L_{e,f}$ by
\[ \pi^L_{e,f}:{\cal E}\ni x\mapsto \<x|e>f\in {\cal E}. \]
We set \emph{${\cal F}_{\cal A}({\cal
E})$}$:=\operatorname{span}\{\pi^L_{e,f}\mid e,f\in {\cal
E}\}$ and call this the set of
 \emph{$\cal A$-finite operators}. The set
 \emph{${\cal K}_{\cal A}({\cal E})$} of \emph{$\cal A$-compact
operators} is
the closure of   ${\cal F}_{\cal A}({\cal E})$ in ${\cal L}_{\cal
A}({\cal E})$.
\end{definition}

\begin{remark}[basic properties]
\begin{enumerate}
\item If ${\cal E}$ is a pre-Hilbert ${\cal A}$-module, $e\in {\cal E}$,
then Definition \ref{definition:HRM}.\ref{enum:fe=ef*} implies $\<e|e>\in{\cal
A}_{sa}$ so that the condition $\<e|e>\geq0$ in
\ref{definition:HRM}.\ref{enum:e=0} makes sense indeed.
\item If ${\cal E}$ is a pre-Hilbert ${\cal A}$-module, $e,f\in {\cal
E},a\in{\cal A}$ then we have:
\[ \<ea|f> =  \<f|ea>^* =  (\<f|e>a)^* =  a^*\<f|e>^* = a^*\<e|f> \]
I.e.\ we have  $\C$- and $\cal A$-sesqui-linearity.
\item The $\C$-sesqui-linearity follows for unital $\cal A$ from the
$\cal A$-sesqui-linearity.
\item For $e,f\in {\cal E}$ we have $\pi_{e,f}^*=\pi_{f,e}$ so that
indeed ${\cal F}_{\cal A}({\cal E})\subset {\cal L}_{\cal A}({\cal
E})$.
\item ${\cal L}_{\cal A}({\cal E})$ and ${\cal K}_{\cal A}({\cal E})$
are \CS-algebras, the former is the multiplier algebra of the latter
\cite[see e.g.\ ][chapter 15]{Weg:KTCAFA}.
\item Everything analogous for left Hilbert modules.
\item ${\cal A}\times{\cal A}\ni(a,b)\mapsto a^*b\in{\cal A}$ together
with  multiplication of ${\cal A}$ on ${\cal A}$ on the right gives the
standard Hilbert ${\cal A}$-module structure on ${\cal A}$.
\item ${\cal A}\times{\cal A}\ni(a,b)\mapsto ab^*\in{\cal A}$ together
with multiplication of ${\cal A}$ on ${\cal A}$ on the left gives the
standard left Hilbert ${\cal A}$-module structure on ${\cal A}$.
\end{enumerate}
\end{remark}

\begin{definition}[free and projective Hilbert modules]
A  Hilbert ${\cal A}$-module is called \emph{free} if it is a free
module over  $\cal A$. It is called \emph{projective} if it is a direct
summand of a free module.
\end{definition}

\begin{lemma}[left and right Hilbert modules]\label{lemma:HLRM}
Let $\cal A$ be a \CS-algebra and $({\cal E},\<\cdot|\cdot>)$ a left
pre-Hilbert $\cal A$-module over $\cal A$. Then
\begin{equation}\begin{split} \<\cdot|\cdot>^{{\cal E}^{op}}:{\cal
E}^{op}\times {\cal E}^{op}&\rightarrow {\cal A}^{op} \\
 (e^{op},f^{op})&\mapsto (\<f|e>)^{op}
\end{split}\end{equation}
defines on ${\cal E}={\cal E}^{op}$ (equality as vector spaces) the structure
of a pre-Hilbert ${\cal A}^{op}$-module.

Furthermore, for a left
Hilbert $\cal A$-module $({\cal E},\<\cdot|\cdot>)$ we have ${\cal
F}_{\cal A}({\cal E})\simeq{\cal F}_{{\cal A}^{op}}({\cal E}^{op})$ and
therefore ${\cal K}_{\cal A}({\cal E})\simeq{\cal K}_{{\cal A}^{op}}({\cal
E}^{op})$ and ${\cal L}_{\cal A}({\cal E})\simeq{\cal L}_{{\cal
A}^{op}}({\cal E}^{op})$.
\end{lemma}
\begin{proof}
It is well known that right $\cal A$-modules ${\cal E}$ and  left ${\cal A}^{op}$-%
modules ${\cal E}^{op}$ are in one-to-one correspondence. So we just have to verify
the corresponding Hilbert module structures:
Let $e^{op},f^{op}\in {\cal E}^{op},a^{op}\in{\cal A}^{op}$. We denote by
$a^{op}$ and $a$ corresponding\footnote{${\cal A}^{op}$ and ${\cal A}$ are
identical as Banach spaces, and in this sense $a^{op}=a$.} elements in
${\cal A}^{op}$ resp.\ ${\cal A}$. then
\begin{align*}
\<e^{op}|f^{op}a^{op}>^{{\cal E}^{op}} &=
\left(\<e^{op}|(af)^{op}>\right)^{op} \\
  &= \<af|e> \\
  &= a\<f|e> \\
  &= \left(\<f|e>\right)^{op}a^{op} \\
  &= \<e^{op}|f^{op}>^{{\cal E}^{op}}a^{op}.
\end{align*}

Since ${\cal E}={\cal E}^{op}$ as Banach space we have ${\cal L}({\cal
E})\simeq{\cal L}({\cal E}^{op})$. Furthermore, for $e,f,x\in {\cal
E}$
\begin{align*}
\pi_{e,f}(x) &= e\<f|x> \\
 &= (\<f|x>)^{op}e^{op} \\
 &= \<x^{op}|f^{op}>^{{\cal E}^{op}}e^{op} \\
 &= \pi^L_{f^{op},e^{op}},
\end{align*}
so that ${\cal F}_{\cal A}({\cal E})$ and ${\cal F}_{{\cal
A}^{op}}({\cal E}^{op})$ are isomorphic, and so are the corresponding
closures and multiplier algebras.
\end{proof}

\begin{remark}[standard module]
For the standard Hilbert ${\cal A}$-module structure on ${\cal A}$ it is
well known that ${\cal F}_{\cal A}({\cal A})={\cal A}$, ${\cal K}_{\cal
A}({\cal A})={\cal A}$ and ${\cal L}_{\cal A}({\cal A})={\cal M}({\cal
A})$.
If we denote by ${\cal A}^L$ the standard left Hilbert ${\cal A}$-%
module then Lemma~\ref{lemma:HLRM} shows: ${\cal F}_{\cal A}({\cal
A}^L)\simeq{\cal F}_{{\cal A}^{op}}({\cal A}^{op})={\cal A}^{op}$.
\end{remark}

\section{GNS representation for Hilbert \CS-modules}\label{sec:GDHCM}
Let $\cal A$ be a \CS-algebra, $\tau$ a state on $\cal A$ and
${\cal E}$ a Hilbert $\cal A$-module.
Analogously to the well know GNS representation of  Banach *-algebras
we define a  scalar product\footnote{For left  Hilbert modules the
 scalar product must be reversed so that one gets complex linearity on
the correct entry.} on ${\cal E}$ by
\begin{equation}
\<x|y>_\tau := \tau(\<x|y>_{\cal E})\text{ for }x,y\in {\cal E}.
\end{equation}
$N_\tau:=\{x\in {\cal E}\mid \<x|x>_\tau=0\}$ is the corresponding null space.
Then the GNS representation space  ${\cal E}_\tau$ is given by the
completion of  ${\cal E}/N_\tau$ with respect to
$\<\cdot|\cdot>_\tau$. $L\in{\cal L}_{\cal A}({\cal E})$ acts
continuously on $x\in {\cal E}_\tau$ because
\begin{align*}
\|Lx\|_\tau^2 &= \<Lx|Lx>_\tau \\
 &= \tau(\<Lx|Lx>_{\cal E}) \\
 &= \tau(\<x|L^*Lx>_{\cal E}) \\
 &\leq \tau(\<x|x>_{\cal E})\|L^*L\| \\
 &= \|x\|_\tau^2 \|L\|^2.
\end{align*}
Thus we have a *-representation of ${\cal L}_{\cal A}({\cal E})$ in
${\cal L}({\cal E}_\tau)$.

If  $\tau$ is faithful then $N_\tau=0$ so that the representation is faithful.

If ${\cal E}={\cal A}$ with $\<a|b>_{\cal E}=a^*b$ is the standard
Hilbert $\cal A$-module then we get back the usual GNS representation of the
multiplier algebra ${\cal L}_{\cal A}({\cal A})={\cal M}({\cal A})$ and,
by restriction, the GNS representation of ${\cal K}_{\cal A}({\cal
A})={\cal A}$.

\input \jobname.bbl
\end{document}